\documentclass[aerospace,article,accept,moreauthors]{Definitions/mdpi}

\externalbibliography{yes}
\renewcommand{\hl}[1]{#1} % arXiv: disable MDPI proof highlighting

\usepackage{algorithm}
 \usepackage{algorithmicx}
 \usepackage{algpseudocode}
\firstpage{1}
\pubvolume{1}
\issuenum{1}
\articlenumber{0}
\pubyear{2026}
\copyrightyear{2026}
\externaleditor{Firstname Lastname}
\datereceived{23 August 2026}
\daterevised{28 September 2026}
\dateaccepted{28 September 2026}
\datepublished{ }

\Title{A Probabilistic Trajectory and Dispersion Study of Martian Tumbleweed Rover Swarms in Global Dust Storm Conditions}

\Author{\hl{Emma C. Belhadfa} %MDPI: Please carefully check the accuracy of names and affiliations.
 $^{1,2,}$*\orcidA{} and \hl{Gabriele M. T. D'Eleuterio} %MDPI: The names highlighted is different from the one submitted online at susy.mdpi.com. Please confirm which is correct. This is correct here.
 $^{2}$}

\AuthorNames{Emma C. Belhadfa and Gabriele M. T. D'Eleuterio}

\address{%
$^{1}$ \quad \hl{Clarendon Laboratory,} %MDPI: We arranged the authors’ address information from subordinate to superior. Please check and confirm this modification. Confirmed.
 Department of Physics, University of Oxford, \mbox{Oxford OX1 3PU, UK}\\
$^{2}$ \quad Institute for Aerospace Studies, University of Toronto, \mbox{Toronto, ON M3H 5T6, Canada}; \hl{gabriele.deleuterio@utoronto.ca} %MDPI: We added these email address here according to which submitted online at susy.mdpi.com. Please confirm. Confirmed.
}

\corres{Correspondence: emma.belhadfa@physics.ox.ac.uk}

\abstract{Global Dust Storms (GDSs) dominate the Martian climate, yet the mechanisms of dust lifting, transport, and deposition remain poorly constrained because in situ observations are limited: landers are stationary, and their Radioisotope Thermoelectric Generators (RTGs) contaminate local measurements. Our solution is a swarm of lightweight, wind-propelled spherical rovers---Dust Rovers (DRs)---that follow natural wind patterns to collect high-spatial-resolution in situ dust data during storms. Here, we quantify the dispersion and communication connectivity of this robotic architecture. Using 625 wind states calculated using the Mars Climate Database (MCD) for the MY34 storm at the Curiosity site, we build a bivariate-normal model of the near-surface wind velocity and characterise its diurnal structure. Rover trajectories and swarm dispersion are propagated with three integrators: the Euler and fourth-order Runge--Kutta schemes for the deterministic dynamics, and a stochastic Ornstein--Uhlenbeck formulation that simulates turbulent gust forcing. Dispersion is controlled by the unmeasured gust correlation time: with a static rolling threshold enforced, the white-gust case ($\tau_g = \Delta\text{t} = 1$~s) provides a two-rover separation below 1~m over an hour (0.44~km over a sol), whereas correlation times of 10--60~s yield 6--25~m over an hour (1.4--3.3~km over a sol). We demonstrate that a 20-rover swarm needs only a 1.6--4.1~km per-rover range to remain connected over a sol, and stays connected across the plausible envelope---for gust amplitudes up to twice the nominal value with $\tau_g\le60$~s, and at the nominal amplitude up to $\tau_g\approx150$~s.}


\begin{document}

%=================================================================
\section{Introduction}
\label{intro}
The interannual variability of Mars' climate is dominated by aerosol forcing, in~particular airborne dust. Approximately three times in 10 Mars Years (MYs), a~Global Dust Storm (GDS) occurs, suspending and transporting fine dust across the entire planet for weeks or months at a time. Despite their central role in material transport and atmospheric processes, the~onset, intensity, and~evolution of GDSs remain difficult to forecast because the mechanisms that trigger and sustain them are poorly constrained by observation~\citep{barnes2020,wolkenberg2020}. These storms also pose an existential hazard to surface missions. During~the MY34 global storm, the~solar-powered Opportunity rover was starved of power and lost~\citep{staab2020}, whereas the nuclear-powered Curiosity rover survived and returned rare in situ measurements of storm dynamics~\citep{chenchen2019,lemmon2019}. Curiosity's data, however, were acquired from a single fixed location, and~its Radioisotope Thermoelectric Generator (RTG) interfered with the dust environment it sought to measure. The~result is a persistent observational gap: no mission to date has been designed to move with, and~thrive within, a~Martian dust storm while sampling it densely in~space.

Closing this gap requires a platform that is (i) mobile in the strong, variable winds of a storm, (ii) spatially distributed, and~(iii) minimally invasive to the dust field being measured. A~robotic swarm---a decentralised, self-organising collective of relatively simple agents~\citep{iglesias2020,brambilla2013}---meets the second and third of these requirements naturally. The~first is met by tumbleweed rovers: lightweight, wind-propelled spheres first proposed in the early 2000s~\citep{antol2003,ylikorpi2005,southard2007,forbes2010} that harness ambient wind for locomotion and therefore follow, rather than fight, the~migration of dust. Because~they carry no active propulsion and no RTG, tumbleweeds disturb the local environment far less than a conventional wheeled rover. They can be manufactured on Earth and stowed for flight in~numbers. 

Previous investigations~\citep{arctic} have deployed similar rovers on Earth, yet the Martian environment reshapes the design problem in ways that do not transfer directly. Foremost is the atmosphere: at roughly 1.6\% of Earth's density \citep{williams_mars_factsheet}, a~given wind speed delivers one to two orders of magnitude less propulsive drag, so a wind-driven vehicle must be far larger and lighter to move---the primary driver behind the 4~m, 20~kg baseline adopted here. Together with Mars' lower gravity, its distinct wind statistics and diurnal cycle, the~absence of any positioning or relay infrastructure, and~the impossibility of recovery or repair, these conditions require that both the vehicle and the swarm's dispersion behaviour be re-derived from Martian first principles rather than inherited from terrestrial~designs.

The platform we analyse is a swarm of wind-propelled vehicles, which we call Dust Rovers (DRs), each carrying six equally spaced photon sensors to measure dust size, shape, and~optical properties in situ. Current models cannot predict the precise starting location or onset time of a GDS~\citep{barnes2020}. Therefore, the~central question for a passive swarm is not whether it can be steered to a storm, but~whether, once deployed, it will remain a coherent, communicating swarm as the wind carries it. This paper answers that question quantitatively. We characterise the near-surface wind during a real storm from modelled climate data, establish the basic mobility of a DR from first principles, and~propagate the resulting rover trajectories and swarm dispersion using both deterministic and stochastic numerical methods. The~stochastic treatment, built on the Ornstein--Uhlenbeck (OU) process, is developed in detail because it is the natural framework for the turbulent, gust-driven forcing that ultimately determines whether the swarm holds~together. 

\textls[-15]{The broader engineering of the mission is not treated here. Vehicle-level aspects, including locomotion, inflation and deployment, power, and~pendulum-based attitude control,~have} been addressed in tumbleweed concept studies, e.g.,~\cite{antol2003,ylikorpi2005, forbes2010}. Entry, descent and landing, thermal survival, telecommunications hardware, autonomous coordination, and~planetary protection are challenges common to any Mars surface mission. Earlier work examined tumbleweed group behaviour only under nominal winds~\cite{southard2007}, and~recent field campaigns have advanced the hardware and begun testing swarm-coordination strategies in Mars-analogue terrain~\cite{kingsnorth2025}---however, neither has quantified the dispersion and connectivity of such a swarm under storm~conditions.

The scope of this study is thus confined to swarm dispersion and connectivity under GDS winds, and its contributions are threefold. We first separate the near-surface storm wind into a swarm-common diurnal mean and a rover-specific gust residual, such that the object of analysis is the cohesion of the swarm rather than the displacement of any single rover. We then represent the gust as a coloured (OU) process with an explicit correlation time $\tau_g$ and, as~this timescale is not constrained by the available data, bound its effect on the dispersion by sweeping it across a plausible range. Finally, we obtain quantitative estimates of the two-rover separation and a network model of the swarm's communication connectivity---to our knowledge the first graph-theory connectivity treatment for wind-driven Martian rover swarms (cf. the mission-concept discussion of \citep{kingsnorth2025})---from which we obtain the critical communication range for connectivity, the~network's robustness to rover loss, and~the storm duration over which Line-Of-Sight (LOS) relay can hold the swarm~together.

%=================================================================
\section{Materials and~Methods}\label{sec:methods}

This section presents the maturity of the tumbleweed concept and the proposed swarm, a~first-principles analysis of DR mobility, the~construction of a multivariate wind model, and~the deterministic and stochastic methods used to propagate rover~trajectories.

\subsection{Platform Background: Tumbleweed Dust~Rovers}

The proposal is deliberately simple: lightweight, spherical rovers pushed by the wind to follow natural weather patterns~\citep{antol2003}. Their passive locomotion and simple structure make them well-suited to mass manufacture and swarm deployment~\citep{iglesias2020}, and~their low mass permits power and thermal solutions that do not rely on an RTG~\citep{ylikorpi2005}. Because~they naturally follow wind patterns, DRs are ideal candidates to track storms and collect data from~within. 

\subsubsection{Science~Traceability}
\label{traceability}
The value of a DR swarm lies in its ability to answer key open science questions about GDSs. No platform is currently able to follow and measure a GDS in situ, across a large spatial region. However, for~the swarm to successfully address these questions, it must remain a connected relay network capable of measuring and transmitting~data.

\textls[-15]{Two lines of prior work approach this question without answering it. Southard~et~al.~\cite{southard2007}} simulated groups of tumbleweed rovers, but~under nominal winds and without a communication model, so neither a storm-driven dispersion rate nor the resulting link requirement was obtained. Kingsnorth~et~al.~\cite{kingsnorth2025} advanced the hardware and begun testing swarm-coordination strategies in Mars-analogue terrain, where the atmosphere is two orders of magnitude denser and relay infrastructure is available, so neither the mobility threshold nor the dispersion statistics transfer. Therefore, neither work establishes whether a passive swarm survives a GDS as a connected~network.

Table~\ref{tab:trace} traces the open science questions to the need for swarm connectivity. Every objective that motivates the concept
resolves, in~its final column, to~the same requirement: that the dispersing swarm
stay in communication. Therefore, the~feasibility of the whole concept of a tumbleweed swarm is gated by the connectivity established in Section~\ref{sec:comm}, motivating the need for the analysis and simulations conducted in this~study.

\begin{table}[H]
\caption{\hl{Science} %MDPI: 1. Please note that changes to the position/size of figures or tables may occur during the production stage. 2. Please confirm the format changes of all tables, including alignments, border lines... Confirmed
 traceability for a DR swarm. Key objectives defined across the literature are outlined, with~current gaps described. We show that the ability of a DR swarm---or any other multi-vehicle platform---to deliver on its science objectives hinges on the connectivity analysis conducted here
(Section~\ref{sec:comm}).\label{tab:trace}}

\begin{adjustwidth}{-\extralength}{0cm}
%\centering %% If there is a figure in wide page, please release command \centering, for Table, ``\textwidth" should be ``\fulllength"
\begin{tabularx}{\fulllength}{LLL}
\toprule
\textbf{Science Objective} & \textbf{Gap Under GDS Conditions} & \textbf{Dependence on Swarm Connectivity}\\
\midrule
Spatial structure of near-surface dust lifting and loading through storm onset and growth~\citep{rs16142613} & Landers fix a single point~\citep{chenchen2019,lemmon2019}; orbiters lose the surface beneath the dust column; storm timing and location are unpredictable~\citep{barnes2020,wolkenberg2020} & A spatial map exists only if the distributed samples are geolocated and relayed as one dataset---requiring a connected swarm (Section~\ref{sec:comm})\\

\bottomrule
\end{tabularx} % if “}” necessary
\end{adjustwidth}
\end{table}

\begin{table}[H]\ContinuedFloat

\caption{\textit{Cont.}}
\begin{adjustwidth}{-\extralength}{0cm}
%\centering

\begin{tabularx}{\fulllength}{LLL}
\toprule
\textbf{Science Objective} & \textbf{Gap Under GDS Conditions} & \textbf{Dependence on Swarm Connectivity}\\
\midrule
In situ particle size distribution and shape across the storm, to~constrain particle-scattering models~\citep{PORTO2024114778} & Size and shape retrievals to date are column-integrated (orbital) or single-point (Curiosity cameras)~\citep{chenchen2019}; no distributed in situ microphysics exists for a GDS & Comparison across the swarm needs simultaneous multi-point returns as it disperses, motivating hop-by-hop~relay\\
\midrule
Dust optical properties (opacity, scattering) mapped across the storm for radiative-transfer modelling~\citep{Martikainen_2024} & Optical depth is constrained only from orbit or a single lander; dark storm interiors are poorly sampled~\citep{lemmon2019} & The scientific product is the spatial field, realised only if the connected swarm returns its~data\\
\midrule
Near-surface dust transport, using the rovers as Lagrangian tracers of the storm wind field~\citep{Jackson_2025} & No in situ Lagrangian tracers have flown in a GDS; fixed stations yield Eulerian point data only & Trajectories inform only while positions are continuously relayed; lost connectivity truncates the transport~record\\
\bottomrule
\end{tabularx}
\end{adjustwidth}
\end{table}

\subsubsection{Vehicle and Payload~Configuration}

\begin{figure}[H]

\begin{adjustwidth}{-\extralength}{0cm}
\centering %% If there is a figure in wide page, please release command \centering, for Table, ``\textwidth" should be ``\fulllength"
\includegraphics[width=0.95\fulllength]{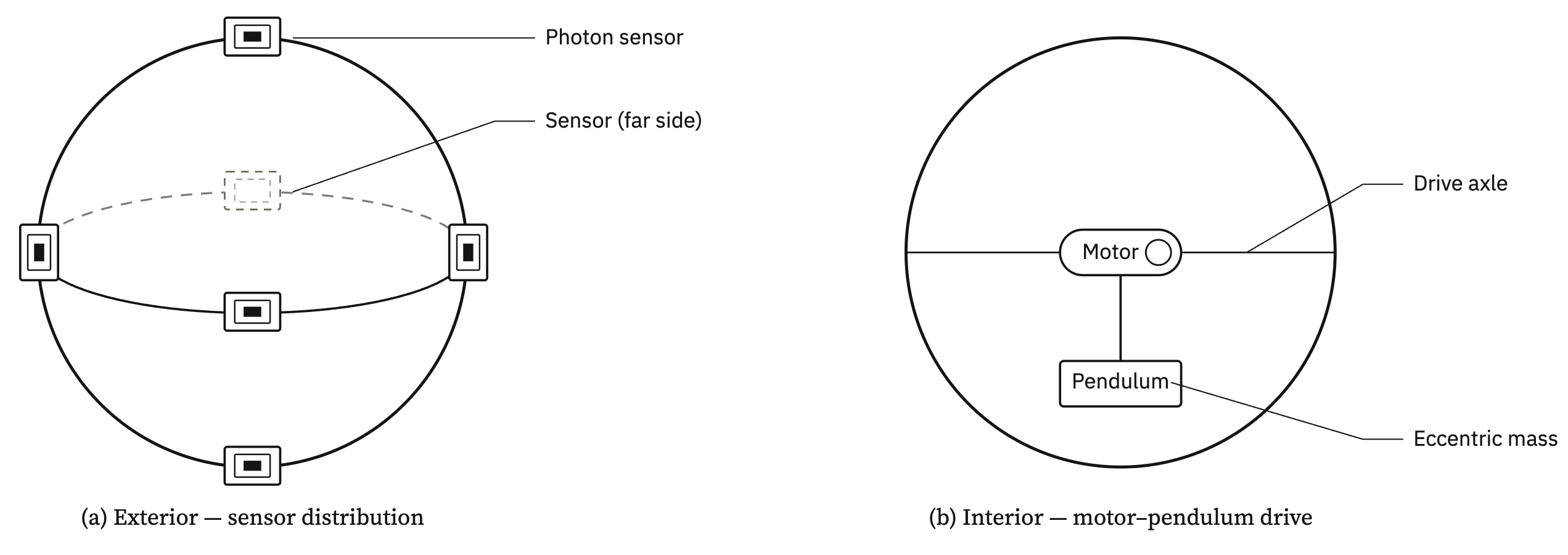}
\end{adjustwidth}
\caption{\hl{The} %MDPI: Please note that each subfigure should be described separately in the caption., e.g., (a) xxx; (b) xxx.
 proposed Dust Rover (DR): external view (\textbf{a}) and internal motor--pendulum and payload layout (\textbf{b}), adapted from~\cite{forbes2010}.\label{fig:drsketch}}
\end{figure}

Every in situ mission flown to Mars has used a traditional rover or helicopter architecture, and~few have operated through a GDS. To~collect the optical dust data needed to constrain particle-scattering models even in dark, extreme storm conditions, a~mission must instead be able to image during the storm and from many points at once. A~DR carries six equally spaced photon sensors (Figure~\ref{fig:drsketch}). Previous quasi-static analyses established that such a vehicle can overcome surface friction and travel long distances with no active propulsion~\citep{antol2003}, and~various surface treatments---fabric sails, turbine-like drag surfaces---and stowage schemes that fold or compress the vehicle for transport have been proposed to increase the number carried per unit volume~\citep{ylikorpi2005,southard2007}. Critically, the~DR contains an internal motor--pendulum system~\citep{forbes2010} capable of generating and storing energy for small attitude corrections, ensuring the sensors remain exposed to the environment. The~vehicle is otherwise propelled solely by the wind: the driving forces are the aerodynamic thrust of the wind and the friction of contact with the surface~\citep{antol2003}.

\subsubsection{Swarm~Architecture}

\begin{figure}[H]

\begin{adjustwidth}{-\extralength}{0cm}
\centering %% If there is a figure in wide page, please release command \centering, for Table, ``\textwidth" should be ``\fulllength"
\includegraphics[width=0.95\fulllength]{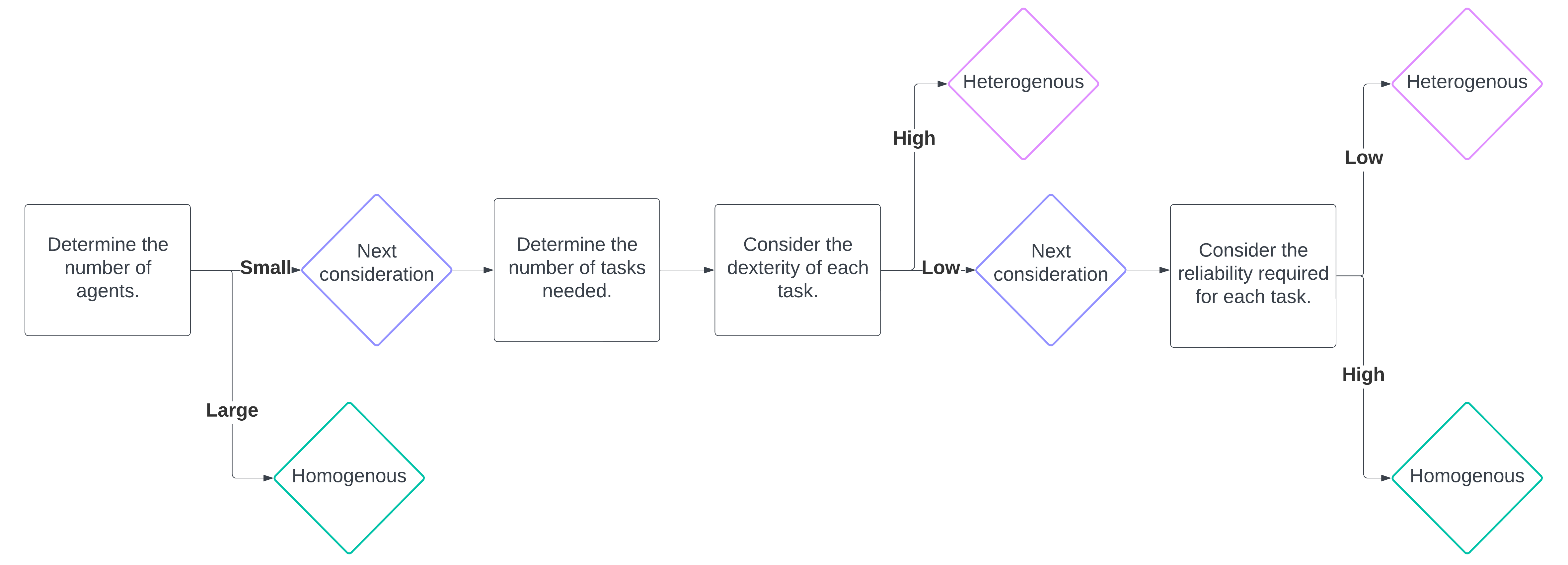}
\end{adjustwidth}
\caption{Trade flow for the swarm architecture decision. Rectangular boxes represent design considerations and questions posed to inform architecture selection. This diagram walks through the decisions made to select a homogeneous mission~architecture. \label{fig:archtrade}}
\end{figure}

Although a swarm may in principle be heterogeneous, the~DR mission calls for a homogeneous architecture (Figure~\ref{fig:archtrade}). The~number of agents is constrained by launch mass and stowed volume and is assumed to be modest (<$20$)~\citep{antol2003}. The~three required \mbox{tasks---travelling} autonomously as a group, collecting in situ dust data, and~communicating with the nearest neighbour---demand little dexterity, since sensing is passive. However, failure of any one task ends the mission, so high reliability is paramount: a homogeneous swarm ensures that no single DR is mission-critical, while simplifying design and manufacture and lowering early phase cost~\citep{brambilla2013}.

\subsubsection{Manufacturing and~Storage}
Previous investigations~\citep{antol2003, ylikorpi2005} have proposed various manufacturing and storage techniques. The~most common, and~best suited to this swarm, would be to manufacture the swarm on Earth and store them flat during transit. Upon~insertion on Mars' surface, the~tumbleweeds would inflate to their full deployed diameter, unfolding from a compact stowed volume into load-bearing spheres. A~particularly attractive variant is the ``Wedges'' concept~\citep{antol2003}, in~which the sphere is formed from several independent inflatable wedge sections arranged around a central instrument package. This format maximises the apparent diameter for a given stowed volume while providing redundancy, since the loss of a single section need not disable a vehicle rolling over the sharp, jagged Martian~terrain.

\subsection{Dust Rover Dynamics and~Mobility}\label{sec:mobility}

Before modelling the swarm statistically, we establish that a single DR can move and negotiate terrain under Martian conditions. The~relevant atmospheric and vehicle parameters are listed in Table~\ref{tab:params}. Following prior tumbleweed studies~\citep{antol2003,ylikorpi2005}, we model the DR as a smooth, uniform sphere of mass $m$ and radius $r$, beginning at rest and driven in the direction of the wind, with~drag distributed over its cross-section $A=\pi r^2$. The~drag resultant is taken to act through the geometric centre, at~height $r$ above the contact point; this fixes the moment arms used in Equations \eqref{eq1} and \eqref{eq:obstacle}.

\begin{table}[H]
\caption{Atmospheric and dust-rover parameters used throughout this~study.\label{tab:params}}

\begin{tabularx}{\textwidth}{lccL}
\toprule
\textbf{Parameter} & \textbf{Symbol} & \textbf{Value} & \textbf{Notes}\\
\midrule
Gravitational acceleration~~~~~ & $g$ & $3.71~\mathrm{m\,s^{-2}}$ & Mars~surface\\
Atmospheric density & $\rho$ & $0.020~\mathrm{kg\,m^{-3}}$ & Near-surface mean; $\sim$1.6\% of Earth \citep{williams_mars_factsheet}\\
Drag coefficient & $C_d$ & $0.4$ & Sphere~\citep{antol2003}\\

\bottomrule
\end{tabularx} % if “}” necessary
%\end{adjustwidth}
\end{table}

\begin{table}[H]\ContinuedFloat
%\tablesize{\small}
\caption{\textit{Cont.}}
%\begin{adjustwidth}{-\extralength}{0cm}
%\centering

\begin{tabularx}{\textwidth}{lccL}
\toprule
\textbf{Parameter} & \textbf{Symbol} & \textbf{Value} & \textbf{Notes}\\
\midrule
Rolling-resistance coefficient & $C_{rr}$ & $0.06$ & Smooth sphere on regolith~\citep{ylikorpi2005}\\
DR mass & $m$ & $20~\mathrm{kg}$ & Margin for payload and pendulum~\citep{forbes2010}\\
DR diameter & $d$ & $4~\mathrm{m}$ & Baseline; $r=d/2$\\
Cross-sectional area & $A$ & $12.57~\mathrm{m^2}$ & $\pi r^2$\\
\bottomrule
\end{tabularx}
\end{table}

\subsubsection{Minimum Force for~Motion}

A DR at rest begins to roll when the aerodynamic thrust generates a moment about the contact point exceeding the resisting moment of gravity. Rolling resistance is modelled as follows: contact deformation displaces the normal-force resultant a distance $o$ ahead of the geometric contact point, and~the rolling-resistance coefficient is defined by $C_{\mathrm{rr}} = o/r$~\citep{ylikorpi2005}. Taking the resultant drag to act through the centre, the~DR begins to roll when the driving moment exceeds the restoring moment, $f_{\mathrm{drag}} r > mgo$, so the minimum forward \hl{force is} %MDPI: Please carefully check variable formatting (italic, bold, subscript, uppercase, etc.) throughout the manuscript to ensure the formatting is consistent and revise if needed. Confirmed

\begin{equation}
f_{\min} = \frac{m g\, o}{r} = C_{\mathrm{rr}}\, m g .
\label{eq1}
\end{equation}

For the baseline DR, $f_{\min}=4.5~\mathrm{N}$ is independent of diameter and shown by Antol~et~al.~\citep{antol2003} to be readily supplied by the wind. Equivalently, a~DR reaches a wind-relative terminal speed, $v_{\mathrm{rel}}^{\mathrm{term}}$, when drag balances rolling resistance, $\tfrac{1}{2}\rho C_{\mathrm{d}} A\,(v_{\mathrm{rel}}^{\mathrm{term}})^2 = C_{\mathrm{rr}} m g$, yielding $v_{\mathrm{rel}}^{\mathrm{term}} = \sqrt{2 C_{\mathrm{rr}} m g/(\rho C_{\mathrm{d}} A)} = 9.4~\mathrm{m\,s^{-1}}$.

\subsubsection{Obstacle and Slope~Negotiation}

To traverse terrain autonomously, a~DR must surmount rocks and slopes. Balancing the drag moment against the gravitational restoring moment about the top edge of a step obstacle yields a maximum surmountable obstacle height, $h_{\mathrm{obs}}$,
\begin{equation}
h_{\mathrm{obs}} = r\left(1-\sqrt{\frac{(mg)^2}{f_{\mathrm{drag}}^2+(mg)^2}}\right), \qquad f_{\mathrm{drag}}=\tfrac{1}{2}\rho C_{\mathrm{d}} A\, v_{\mathrm{w}}^2 ,
\label{eq:obstacle}
\end{equation}
where $v_{\mathrm{w}}$ is the free-stream wind speed, parallel to the mean~surface. 

The same horizontal wind sets the maximum climbable slope. On~an incline of angle $\theta$, the~drag $f_{\mathrm{drag}}$ projects onto the slope as $f_{\mathrm{drag}}\cos\theta$, whereas the rolling resistance acts tangentially along the slope and is proportional to the normal load, $f_{\mathrm{fric}}=C_{\mathrm{rr}}mg\cos\theta$. The~rover is on the verge of climbing when the up-slope driving force balances the tangential gravity component, $f_{\mathrm{drag}}\cos\theta-C_{\mathrm{rr}}mg\cos\theta = mg\sin\theta$. Hence,
\begin{equation}
    mg\sin\theta_{\mathrm{max}}= (f_\mathrm{drag}-C_{\mathrm{rr}}mg)\cos\theta_\mathrm{max}
\end{equation}
which, on~dividing by $mg\cos\theta_{\max}$, yields
\begin{equation}
\theta_{\max}=\arctan\!\left[\frac{\rho C_d A\, v_{\mathrm{w}}^2}{2mg}-C_{\mathrm{rr}}\right].
\label{eq:slope}
\end{equation}

Both capabilities grow with wind speed and diameter, so a DR is substantially more mobile during storms than in quiescent conditions. The~obstacle capability in Equation~(\ref{eq:obstacle}) uses the free-stream wind appropriate to a stationary rover; a rover already near its rolling speed sees a reduced relative wind and therefore a smaller climbing margin. Rock size-frequency distributions measured for the Mars Exploration Rovers~\citep{golombek2003} show that obstacles below the centimetre scale are common, but~that rocks exceeding it occur at metre-to-decametre spacing at rocky sites, so arrest by rock fields is a credible competing mechanism. We therefore limit the mobility claims to unobstructed~terrain.

\subsection{Wind~Characterisation}\label{sec:wind}

The trajectory of a passively driven DR is set entirely by the wind, so a faithful statistical model of the near-surface wind is the foundation of the~analysis.

\subsubsection{Data~Set}

Curiosity carries no functioning wind sensor, as the Rover Environmental Monitoring Station (REMS) wind sensor was damaged on landing and returned usable data for only about two MYs~\citep{viudez2019}, before~the onset of the MY34 storm. Hence, no in situ wind measurements exist at the site for this event, and~a climatology model is the only practical source of a complete diurnal wind~field.

Here, the~near-surface wind is characterised using the Mars Climate Database v6.1 (MCD)~\citep{forget2017}, a~climatological database compiled from Mars general-circulation-model simulations~\citep{forget1999, Millour2018, Colatis2013}. The~MCD supplies model-derived wind fields: the values used here are synthetic and~are neither measured nor independently validated by instruments at the surface. Therefore, the~dataset is a physically plausible statistical description rather than ground truth, a~limitation we return to in Section~\ref{sec:discussion}. We adopt the MY34 global dust storm, the~most extensively documented recent global event~\citep{lemmon2019}, and~query the database at 2~m above the Curiosity landing site ($4.6^\circ$S, $137.4^\circ$E), taken as a representative low-latitude location where other instruments onboard Curiosity collected data throughout the storm. Three fields were used: the horizontal wind speed ($v$) and the meridional (south--north, $w$) and zonal (west--east, $u$) components, sampled hourly (25 local-time bins) across the active storm period, $160^\circ \le L_s \le 250^\circ$ in 25 steps, yielding $25\times25=625$ wind states. The~three fields are mutually consistent by construction, i.e.,~$v=\sqrt{u^2+w^2}$. This window spans the expansion, peak (i.e., planet encirclement near $L_s \approx 193^\circ$) and decay phases of the MY34 event~\citep{wolkenberg2020}. Because~pre- and post-peak states are included, the~resulting mean velocity describes the storm-period in its entirety, rather than just the peak-storm; we report it as such~throughout.

\subsubsection{Wind-Speed~Distribution}

Wind speed is classically modelled by a two-parameter Weibull distribution~\citep{tuller1984}, with~probability density
\begin{equation}
p_W(v;k,\lambda)=\frac{k}{\lambda}\left(\frac{v}{\lambda}\right)^{k-1}\exp\!\left[-\left(\frac{v}{\lambda}\right)^{k}\right],\qquad v\ge 0,
\end{equation}
where $k$ is the shape parameter and $\lambda$ is the scale~parameter. 

The parameters are estimated by maximum likelihood~\citep{cohen1965}. For~a sample $\{v_i\}_{i=1}^{n}$, the~log-likelihood of the two-parameter Weibull is
\begin{equation}
\ln L_w(k,\lambda)=n\ln k-nk\ln\lambda+(k-1)\sum_{i=1}^{n}\ln v_i
-\sum_{i=1}^{n}\left(\frac{v_i}{\lambda}\right)^{\!k}.
\end{equation}

Setting $\partial\ln L_w/\partial\lambda=0$ provides the scale in closed form,
\begin{equation}
\hat{\lambda}=\left(\frac{1}{n}\sum_{i=1}^{n}v_i^{\,k}\right)^{\!1/k},
\end{equation}
and $\partial\ln L_w/\partial k=0$, after~eliminating $\lambda$, leaves a single equation for the shape,\vspace{+6pt}
\begin{equation}
\frac{\sum_{i=1}^{n}v_i^{\,k}\ln v_i}{\sum_{i=1}^{n}v_i^{\,k}}-\frac{1}{k}
-\frac{1}{n}\sum_{i=1}^{n}\ln v_i=0,
\end{equation}
which is solved numerically for $\hat{k}$, whereupon $\hat{\lambda}$ follows from the closed form above~\citep{cohen1965}. Applied to the 625 wind states, this calculation yields $k=2.72$ and $\lambda=8.52~\mathrm{m\,s^{-1}}$ (storm-period mean speed $7.60~\mathrm{m\,s^{-1}}$; diurnal mean peaks at $9.9~\mathrm{ms}^{-1}$). The~direction must, however, be treated jointly with speed, as~we now~show.

\subsubsection{Joint Structure and Diurnal~Variability}

Modelling horizontal wind speed $v$ and heading $\phi$ as independent variables is convenient but incorrect: the wind direction on Mars is organised by a strong diurnal cycle~\citep{viudez2022}. The~heading is influenced by diurnal tides: a dominant night-time regime near 120--180$^\circ$ with the strongest speeds ($\sim$9--10~m\,s$^{-1}$) and a weaker daytime regime near $-60$--$-90^\circ$. Additional thermal tides, including semi-diurnal wave-2 tides, complicate modelling of all harmonics. Here, we focus on the dominant diurnal tide and treat the system as bimodal. Therefore, speed and heading are statistically dependent, and a single Gaussian placed on the wrapped angle $\phi$ is a poor representation. Rather than force a distribution onto the circular variable, we model the wind's zonal and meridional components, $(u,w)$, directly. This solution avoids the angle-wrapping pathology entirely, embeds the speed--direction dependence in a single covariance, and~yields the wind vector required to force the rover~dynamics.

\subsubsection{Bivariate Normal Component~Model}

\hl{Let} %MDPI: Please confirm if the bold formatting of variables is necessary; if not, please remove it. The same below, please check the whole text. Confirmed
 $\mathbf{V}=(u,w)^{\!\top}$ denote the near-surface wind velocity vector, with~$u$ and $w$ its zonal and meridional components. We model its distribution as bivariate normal,
\begin{equation}
\mathbf{V}\sim\mathcal{N}(\boldsymbol{\mu},\boldsymbol{\Sigma}),
\end{equation}
and estimate the mean vector $\boldsymbol{\mu}$ and covariance matrix $\boldsymbol{\Sigma}$ from the 625 MCD wind states as the sample mean and sample~covariance.

This choice is supported by the central limit theorem---the aggregate of many independent atmospheric perturbations tends to normality~\citep{gut2009}---and by the analytic tractability of the Gaussian, whose dependence structure is fully captured by $\boldsymbol{\Sigma}$ without recourse to copulas~\citep{krupskii2013}, which are computationally expensive. A~single bivariate Weibull would require such a copula construction, which lies outside our scope. Samples are drawn efficiently through the Cholesky factor $\mathbf{L}$ of $\boldsymbol{\Sigma}=\mathbf{L}\mathbf{L}^\top$, via $\mathbf{V}=\boldsymbol{\mu}+\mathbf{L}\mathbf{z}$ with $\mathbf{z}\sim\mathcal{N}(\mathbf{0},\mathbf{I}_2)$.

The bivariate normal above pools all 625 states and so blends the diurnal cycle with the turbulent gusts. To~isolate the fast, rover-specific gusts that drive dispersion from the slow, swarm-common diurnal trend, we bin the same 625 states by local hour. The~states form a $25\times25$ grid in local time and solar longitude, so each of the 25 local-hour bins holds $n_h=25$ states ($\sum_h n_h = 625$). For~bin $h$ we compute---exactly as for $\boldsymbol{\mu}$ and $\boldsymbol{\Sigma}$ above, but~per bin---the sample mean $\bar{\mathbf{V}}_h$; together, these form the diurnal mean field. The~gust residual is the deviation of each state from its own hourly mean,
\begin{equation}
    \boldsymbol{\xi}(t) = \mathbf{V}(t) - \bar{\mathbf{V}}_{h(t)},
\end{equation}
which therefore has zero sample mean within every bin, and~hence zero mean overall, by~construction; only its shape is assumed, namely Gaussian with the pooled residual covariance $\boldsymbol{\Sigma}_g=\mathrm{cov}(\boldsymbol{\xi})$. The~wind is then generated, rather than decomposed, as~\begin{equation}
    \mathbf{V}(t) = \bar{\mathbf{V}}_{h(t)} + \boldsymbol{\xi}(t),
    \qquad
    \boldsymbol{\xi} \sim \mathcal{N}(\mathbf{0}, \boldsymbol{\Sigma}_{g}),
    \label{eq:decomposition}
\end{equation}
with $\bar{\mathbf{V}}_{h(t)}$ the mean at the local hour $h(t)$ of time $t$. The~two descriptions are consistent views of the same data: averaging the hourly means recovers the global mean, $\boldsymbol{\mu}=\tfrac{1}{25}\sum_h\bar{\mathbf{V}}_h$, and~the total covariance separates into between-hour (diurnal) and within-hour (gust) parts,
$\boldsymbol{\Sigma}=\mathrm{cov}(\bar{\mathbf{V}}_h)+\boldsymbol{\Sigma}_g$, so
$\boldsymbol{\Sigma}_g$ is the smaller, within-hour component of $\boldsymbol{\Sigma}$.
Equation~\eqref{eq:decomposition} specifies the marginal distribution of the gust at a given instant; its temporal correlation structure is left open here and is specified in Section~\ref{sec:trajmethods}.

\subsection{Trajectory~Models}\label{sec:trajmethods}

With a probabilistic wind model in hand, we propagate the two-dimensional rover position $\mathbf{p}=(x,y)$ from the starting position at Curiosity's landing site, $\mathbf{p}_0=(0,0)$, over~an hour ($\Delta t=1$~s, $N=3600$ steps) and a full sol ($N=88{,}800$). Writing the rover velocity as $\dot{\mathbf p}$, Newton's second law provides the deterministic force per unit mass:
\begin{equation}
\mathbf{F}(\mathbf{V},\dot{\mathbf p})=\frac{1}{m}\!\left[\tfrac12\rho C_dA\,
\lVert\mathbf{V}-\dot{\mathbf p}\rVert(\mathbf{V}-\dot{\mathbf p})-C_{rr}mg\,
\frac{\dot{\mathbf p}}{\lVert\dot{\mathbf p}\rVert}\right],
\label{eq:F}
\end{equation}
so that the rover state $(\mathbf p,\dot{\mathbf p})$ evolves as a second order equation,
\begin{equation}
\ddot{\mathbf p}=\mathbf F(\mathbf V,\dot{\mathbf p}).
\end{equation}

The rolling resistance direction, $\dot{\mathbf p}/{\lVert\dot{\mathbf p}\rVert}$, is undefined at zero velocity, so the dynamics are integrated with an explicit rest/motion state: a resting rover stays at rest until \mbox{$\lVert F_\mathrm{drag} \rVert > f_\mathrm{min}$}, after~which static friction opposes the drag direction. A~moving rover feels kinetic rolling resistance $f_\mathrm{min}$ opposing its velocity, with~the per-step friction impulse capped at the current speed (the numerical regularisation replacing $\dot{\mathbf p}/{\lVert\dot{\mathbf p}\rVert}$) so it can, at~most, return the rover to rest, at~which point it stays if the drag is~sub-threshold. 

We integrate this system three ways: two deterministic integrators applied as Monte-Carlo ensembles, and~one stochastic integrator based on an OU formulation. Our goal is to compare how considering stochastic effects impacts the resulting dispersion and trajectories, informing the ways in which stochastic considerations (i.e., coloured versus white noise) improve upon traditional, deterministic~methods.

All simulations were implemented in \hl{Python 3.16} %MDPI: Please state which version of the software was used. Added
 (NumPy/SciPy) and executed on a single Apple M3 core; ensembles are fully vectorised over rovers. The~code can be found on the corresponding author's \hl{GitHub} \citep{Belhadfa2026-fq}. Because~each Monte-Carlo member samples the distribution of a single rover's displacement, swarm dispersion is quantified by the distance between two independently drawn ($\ell=0$) rover endpoints, $D=\lVert\mathbf{p}^{(i)}-\mathbf{p}^{(j)}\rVert$. To~fully quantify the range of outcomes, we report the mean and upper percentiles of $D$, which are independent of ensemble size (unlike the extremal separation over $n$ trials).

 Connectivity is a property of the whole swarm rather than of a single pair, so we model the $N$-rover swarm as a time-varying unit-disk graph: the rovers are nodes, an~edge joins any two separated by less than the communication range $r_c$, and~the swarm is connected when the graph is a single component, so that data reach every rover by multi-hop relay. For~each of $R=2000$ Monte-Carlo swarms, we draw $N$ trajectories from the ensemble and evaluate the graph at every recorded step ($600$~s). Two families of metrics follow. From~the end-of-sol (most-dispersed) state, we report the critical range $r_c^\star$---the smallest $r_c$ that connects the swarm, equal to the longest edge of its Euclidean minimum spanning tree~\citep{penrose2003} and reported as the $95$th percentile over the $2000$ swarms---and the algebraic connectivity $\lambda_2$, the~second-smallest Laplacian eigenvalue~\citep{fiedler1973} (\hl{Table}%MDPI: The tables should be mentioned in order. Now, Table 5 is cited before Table 3. Please adjust the order and make sure the tables appears after the first mention. You can move them around if necessary. I feel as though the current order is what was intended.
~\ref{tab:net}). The~time-resolved evaluation provides the fraction of the sol for which the swarm is connected and the outage statistics (Section~\ref{sec:netresults}).

\subsubsection{Deterministic Integration: Euler and~Runge--Kutta}
\label{sec:determ}
The explicit Euler scheme~\citep{hemami1982} advances the state by
\begin{equation}
    \mathbf p_{n+1}=\mathbf p_n+\Delta t\, \dot{\mathbf p}_n,
\end{equation}
\begin{equation}
    \dot{\mathbf p}_{n+1}= \dot{\mathbf p}_n+\Delta t\, \mathbf F(\mathbf V_n,\dot{\mathbf p}_n),
\end{equation}
with $\mathbf{V}_n = \bar{\mathbf{V}}_{h(t_n)} + \boldsymbol{\xi}_n$ and $\boldsymbol{\xi}_n \sim \mathcal{N}(\mathbf{0}, \boldsymbol{\Sigma}_g)$ as an~independent gust draw at each step, such that the deterministic schemes use the same mean field and gust covariance as the stochastic scheme of Section~\ref{sec:stoch} and differ from it only in the gust correlation time. Specifically, it treats gusts as uncorrelated (i.e., white) noise.

The fourth-order Runge--Kutta (RK4) scheme evaluates $\mathbf F$ at four sub-points per step and combines them as a weighted average~\citep{cartwright1992}, giving local error $O(\Delta t^5)$ against Euler's $O(\Delta t^2)$. In~both, the~spread of the ensemble estimates the swarm~dispersion.

To enable a direct comparison across integrators, all methods use identical gust histories (i.e., matched random inputs). 

\subsubsection{Stochastic Model: The Ornstein--Uhlenbeck~Process}
\label{sec:stoch}
The deterministic schemes resample the wind as white noise, but~a turbulent gust is not instantaneous; it persists over a finite correlation time. The~natural formalism for such coloured, mean-reverting forcing is a stochastic differential equation driven by a Wiener process. A~standard, two-dimensional Wiener process $\mathbf{W}_t$ is defined by \mbox{$\mathbf{W}_0=\mathbf{0}$}, independent increments, Gaussian increments $\mathbf{W}_t-\mathbf{W}_s\sim\mathcal{N}(\mathbf{0},(t-s)\mathbf{I}_2)$ for $t>s$, and~continuous sample paths~\citep{malliaris1990}. We model the gust as an OU process with correlation time $\tau_g$ and stationary covariance $\boldsymbol{\Sigma}_g=\mathbf L_g\mathbf L_g^{\!\top}$, driven by the Wiener process $\mathbf W_t$,
\begin{equation}\label{eq:gust}
\mathrm d\boldsymbol{\xi}_t=-\frac{1}{\tau_g}\,\boldsymbol{\xi}_t\,\mathrm dt
+\sqrt{\frac{2}{\tau_g}}\,\mathbf L_g\,\mathrm d\mathbf W_t .
\end{equation}

\hl{Each} %MDPI: We changed the indentation format of some paragraphs after formulas, please check the entire paper and confirm. Confirm.
 rover is driven by the total wind $\mathbf V(t)=\bar{\mathbf V}_{h(t)}+\boldsymbol{\xi}_t$, so its state $(\mathbf p,\dot{\mathbf p})$ obeys
\begin{equation}\label{eq:sde}
\mathrm d\mathbf p_t=\dot{\mathbf p}_t\,\mathrm dt,\qquad
\mathrm d \dot{\mathbf p}_t=\mathbf F\!\left(\bar{\mathbf V}_h+\boldsymbol{\xi}_t,\,\dot{\mathbf p}_t\right)\mathrm dt,
\end{equation}
with $\mathbf F$ the force per unit mass of Equation~\eqref{eq:F}. The~correlation time $\tau_g$ is the single parameter that the hourly MCD data cannot constrain; its influence is quantified in Section~\ref{sec:conv}. Setting $\tau_g=\Delta t$ does not produce a fully independent gust: under the exact transition of Equation~\eqref{eq:emgust}, it leaves a lag-one autocorrelation of $e^{-1}\approx0.37$, and~the independent-draw limit is $\tau_g\ll\Delta t$. The~deterministic Euler and RK4 integrators of Section~\ref{sec:determ} instead draw an independent gust $\boldsymbol{\xi}_n\sim\mathcal{N}(\mathbf 0,\boldsymbol{\Sigma}_g)$ at every step (Table~\ref{tab:sep}), whereas the $\tau_g=\Delta t=1$~s row of the OU sweep (Table~\ref{tab:tau}) retains this small residual correlation. This is why the OU white-gust row gives $\sim1$~m over an hour, while the fully independent deterministic schemes give $0.38$--$0.42$~m.

\begin{table}[H]
\caption{\textls[-15]{\hl{Two-rover} %MDPI: We moved Table 3 after its first citation, please confirm. Confirm
 separation $D$ and mean displacement after one hour under
mean storm-period wind ($n=10{,}000$), matched-seed white-gust
($\tau_g=\Delta t=1$~s) integrators on the static-threshold~dynamics}. \label{tab:sep}}
\begin{tabularx}{\textwidth}{Xcccc}
\toprule
\textbf{Method} & \textbf{Runtime [s]} & \textbf{Mean $\bm{D}$ [m]} & \textbf{99th Pct.\ [m]} & \textbf{Mean Displ.\ [m]}\\
\midrule
Euler & 3.6 & 0.42 & 1.35 & 1.30\\
Runge--Kutta (RK4) & 9.0 & 0.38 & 1.24 & 1.35\\
\bottomrule
\end{tabularx}
\end{table}
\vspace{-6pt}

\begin{table}[H]
\caption{\hl{Two-rover} %MDPI: We moved Table 4 after its first citation, please confirm. Confirm
 separation versus gust correlation time $\tau_g$ under the
static-threshold baseline (hour case: mean storm wind; sol case: diurnal cycle).
The white-gust discrete case ($\tau_g=\Delta t=1$~s) is shown for reference; it
is not a general lower bound but the $\tau_g=1$~s member of the sweep, whose
diffusion scales as $\sqrt{\Delta t}$. The~range is extended to $300$~s to
bracket the loss of one-sol connectivity: the sol $99$th percentile crosses the
$7.4$~km horizon near $\tau_g\approx30$--$35$~s and the sol mean near
$\tau_g\approx300$~s. Hour mean values are rounded to the nearest metre. \label{tab:tau}}
\begin{tabularx}{\textwidth}{cCCCC}
\toprule
\textbf{$\bm{\tau_g}$ [s]} & \textbf{Hour Mean [m]} & \textbf{Hour 99th [m]} & \textbf{Sol Mean [km]} & \textbf{Sol 99th [km]}\\
\midrule
1   & 1  & 2   & 0.44 & 1.32\\
10  & 6  & 34  & 1.37 & 4.08\\
20  & 12 & 71  & 1.94 & 5.80\\
30  & 17 & 100 & 2.37 & 7.13\\
40  & 20 & 129 & 2.74 & 8.22\\
60  & 25 & 181 & 3.34 & 10.08\\
100 & 31 & 275 & 4.29 & 13.04\\
150 & 35 & 350 & 5.23 & 15.98\\
200 & 37 & 413 & 6.02 & 18.48\\
300 & 39 & 512 & 7.34 & 22.68\\
\bottomrule
\end{tabularx}
\end{table}

Equations~\eqref{eq:gust} and~\eqref{eq:sde} are integrated with the exact OU transition for the gust and an explicit Euler step for the
rover state~\citep{kloeden1992,higham2001}. The~update is therefore
\begin{align}
\boldsymbol{\xi}_{n+1}&=e^{-\Delta t/\tau_g}\,\boldsymbol{\xi}_n
+\sqrt{1-e^{-2\Delta t/\tau_g}}\;\mathbf L_g\mathbf z_n,\qquad
\mathbf z_n\sim\mathcal N(\mathbf 0,\mathbf I_2),\label{eq:emgust}\\
\dot{\mathbf p}_{n+1}&=\dot{\mathbf p}_n+\mathbf F\!\left(\bar{\mathbf V}_h+\boldsymbol{\xi}_n,\dot{\mathbf p}_n\right)\Delta t,\qquad
\mathbf p_{n+1}=\mathbf p_n+\dot{\mathbf p}_{n+1}\,\Delta t.\label{eq:emstate}
\end{align}

Because the gust of Equation~\eqref{eq:emgust} is advanced exactly, its stationary covariance $\boldsymbol{\Sigma}_g$ is independent of $\Delta t$, and the only discretisation error is the first-order Euler step~\eqref{eq:emstate}. The~full procedure is presented in Algorithm \ref{algo}.

\begin{algorithm}[H]
\caption{ Stochastic integration of the DR trajectory ensemble   \label{algo}}

\begin{algorithmic}

\State \begin{enumerate}
\item \textbf{Input:} $\mathbf{p}_0$; drift $\mathbf{F}$; gust factor $\mathbf{L}_g$; correlation time $\tau_g$; step $\Delta t$; steps $N$; ensemble size $n$
\item \textbf{for} $j=1$ to $n$ (rovers, in~parallel) \textbf{do}
\item[] \quad $\mathbf{p}\gets\mathbf{p}_0$,\quad $\dot{\mathbf p}\gets\mathbf{0}$,\quad $\boldsymbol{\xi}\gets\mathbf{0}$
\item[] \quad \textbf{for} $i=0$ to $N-1$ \textbf{do}
\item[] \quad\quad $\bar{\mathbf{V}}\gets$ hourly mean wind at local hour $h(t_i)$
\item[] \quad\quad $\boldsymbol{\xi}\gets e^{-\Delta t/\tau_g}\boldsymbol{\xi}+\sqrt{1-e^{-2\Delta t/\tau_g}}\,\mathbf L_g\mathbf z$,\quad $\mathbf z\sim\mathcal N(\mathbf 0,\mathbf I_2)$
\item[] \quad\quad $\dot{\mathbf p}\gets\dot{\mathbf p}+\mathbf F(\bar{\mathbf V}_h+\boldsymbol{\xi},\dot{\mathbf p})\,\Delta t$;\quad $\mathbf p\gets\mathbf p+\dot{\mathbf p}\,\Delta t$
\item[] \quad \textbf{end for}
\item \textbf{end for}
\item \textbf{Output:} rover endpoints $\{\mathbf{p}^{(j)}\}_{j=1}^{n}$
\end{enumerate}

\end{algorithmic}

\end{algorithm}

%\begin{center}
%\fbox{\begin{minipage}{0.93\linewidth}
%\textbf{\hl{Algorithm 1}.} \\[3pt]
%\label{algo}
%\rule{\linewidth}{0.4pt}
%\begin{enumerate}[leftmargin=2.2em,itemsep=1pt,topsep=3pt]
%\item \textbf{Input:} $\mathbf{p}_0$; drift $\mathbf{F}$; gust factor $\mathbf{L}_g$; correlation time $\tau_g$; step $\Delta t$; steps $N$; ensemble size $n$
%\item \textbf{for} $j=1$ to $n$ (rovers, in~parallel) \textbf{do}
%\item[] \quad $\mathbf{p}\gets\mathbf{p}_0$,\quad $\dot{\mathbf p}\gets\mathbf{0}$,\quad $\boldsymbol{\xi}\gets\mathbf{0}$
%\item[] \quad \textbf{for} $i=0$ to $N-1$ \textbf{do}
%\item[] \quad\quad $\bar{\mathbf{V}}\gets$ hourly mean wind at local hour $h(t_i)$
%\item[] \quad\quad $\boldsymbol{\xi}\gets e^{-\Delta t/\tau_g}\boldsymbol{\xi}+\sqrt{1-e^{-2\Delta t/\tau_g}}\,\mathbf L_g\mathbf z$,\quad $\mathbf z\sim\mathcal N(\mathbf 0,\mathbf I_2)$
%\item[] \quad\quad $\dot{\mathbf p}\gets\dot{\mathbf p}+\mathbf F(\bar{\mathbf V}_h+\boldsymbol{\xi},\dot{\mathbf p})\,\Delta t$;\quad $\mathbf p\gets\mathbf p+\dot{\mathbf p}\,\Delta t$
%\item[] \quad \textbf{end for}
%\item \textbf{end for}
%\item \textbf{Output:} rover endpoints $\{\mathbf{p}^{(j)}\}_{j=1}^{n}$
%\end{enumerate}
%\end{minipage}}
%\end{center}

Rovers close together sample much of the same flow, so we correlate the gust spatially: the forcing covariance between two rovers, \textit{a} and \textit{b}, is $C(\delta_{ab}) \Sigma_g$, where $C(\delta) = \mathrm{exp}(-\delta/\ell)$ and $\delta$ is the rover separation. Co-located rovers therefore share the single-point covariance and the relative forcing vanishes as $\delta \rightarrow 0$. The~correlation length $\ell$ is treated as a sensitivity parameter; the independent-gust case ($\ell=0$) is retained as the conservative bound. In~practice, for~a swarm of $N$ rovers, we build the $N\times N$ spatial-coherence matrix $C_{ab}=\exp(-\delta_{ab}/\ell)$ from the current pairwise separations $\delta_{ab}$, take its Cholesky factor $\mathbf{C}=\mathbf{L}_s\mathbf{L}_s^{\!\top}$, and~apply $\mathbf{L}_s$ to the per-rover white-noise draws before the single-point OU update of Equation~\eqref{eq:emgust} (whose factor $\mathbf{L}_g$ sets the per-rover covariance); $\mathbf{C}$ is recomputed periodically as the rovers separate. The~spatial-correlation results (\hl{Table}%MDPI: The tables should be mentioned in order. Now, Table 6 is cited before Table 5. Please adjust the order and make sure the tables appears after the first mention. Confirm.
~\ref{tab:spatial}, \hl{Figure}%MDPI: The figures should be mentioned in order. Now, Figure 10 is cited before Figure 2. Please adjust the order and make sure the figures appears after the first mention. Confirm
~\ref{spatial_sensitivity}) are therefore produced by joint $N$-rover swarm simulations. By~contrast, the~two-rover statistic $D$ defined above is computed from independently drawn ensemble members and thus assumes independent forcing; it is used only for the $\ell=0$ results and is not applied for $\ell>0$.

%=================================================================
\section{Results}\label{sec:results}

\subsection{Wind~Model}

\begin{figure}[H]

\begin{adjustwidth}{-\extralength}{0cm}
\centering %% If there is a figure in wide page, please release command \centering, for Table, ``\textwidth" should be ``\fulllength"
\begin{minipage}{0.63\textwidth}\centering
\includegraphics[width=\linewidth]{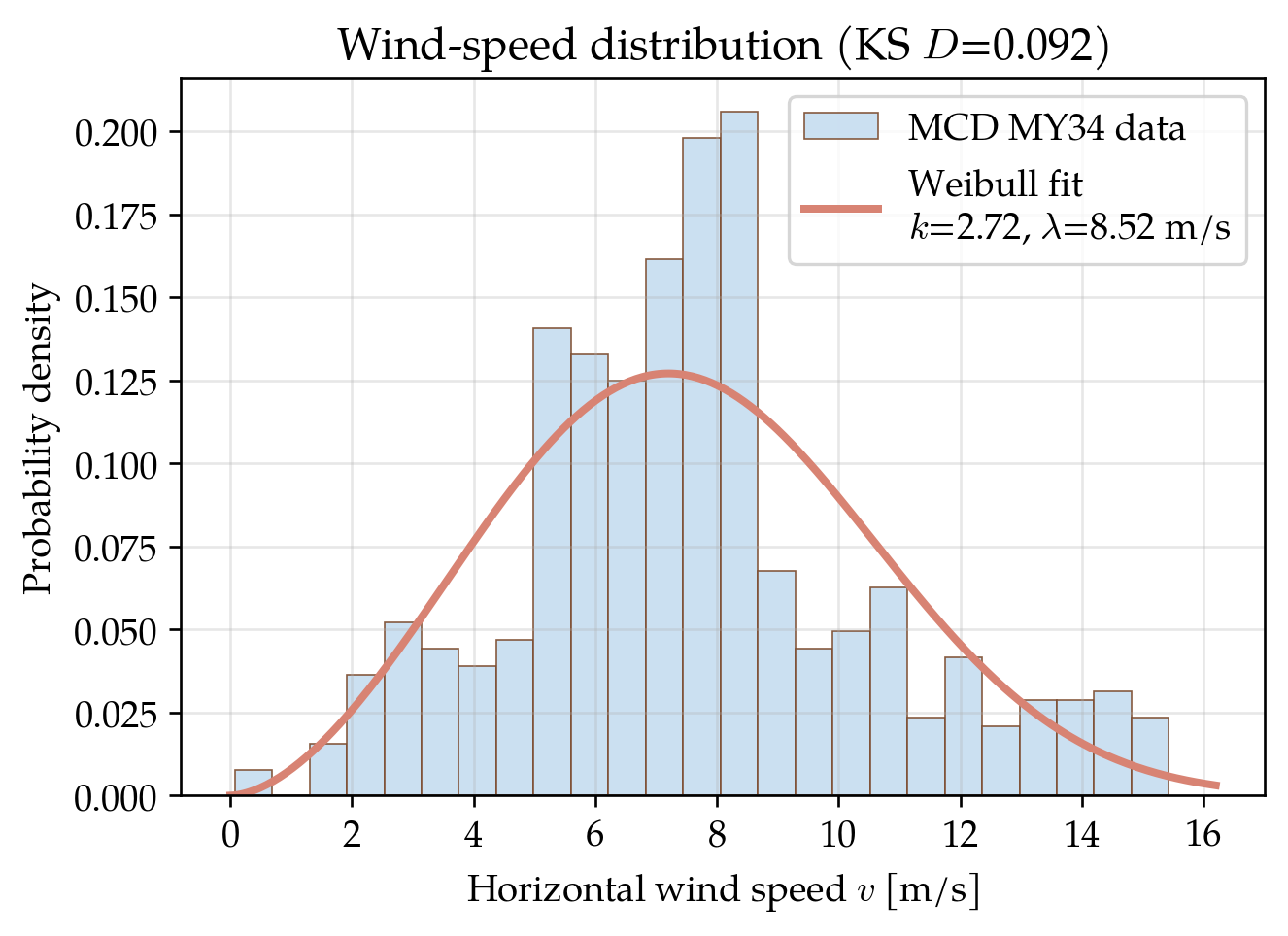}\\(\textbf{a})
\end{minipage}
\begin{minipage}{0.63\textwidth}\centering
\includegraphics[width=\linewidth]{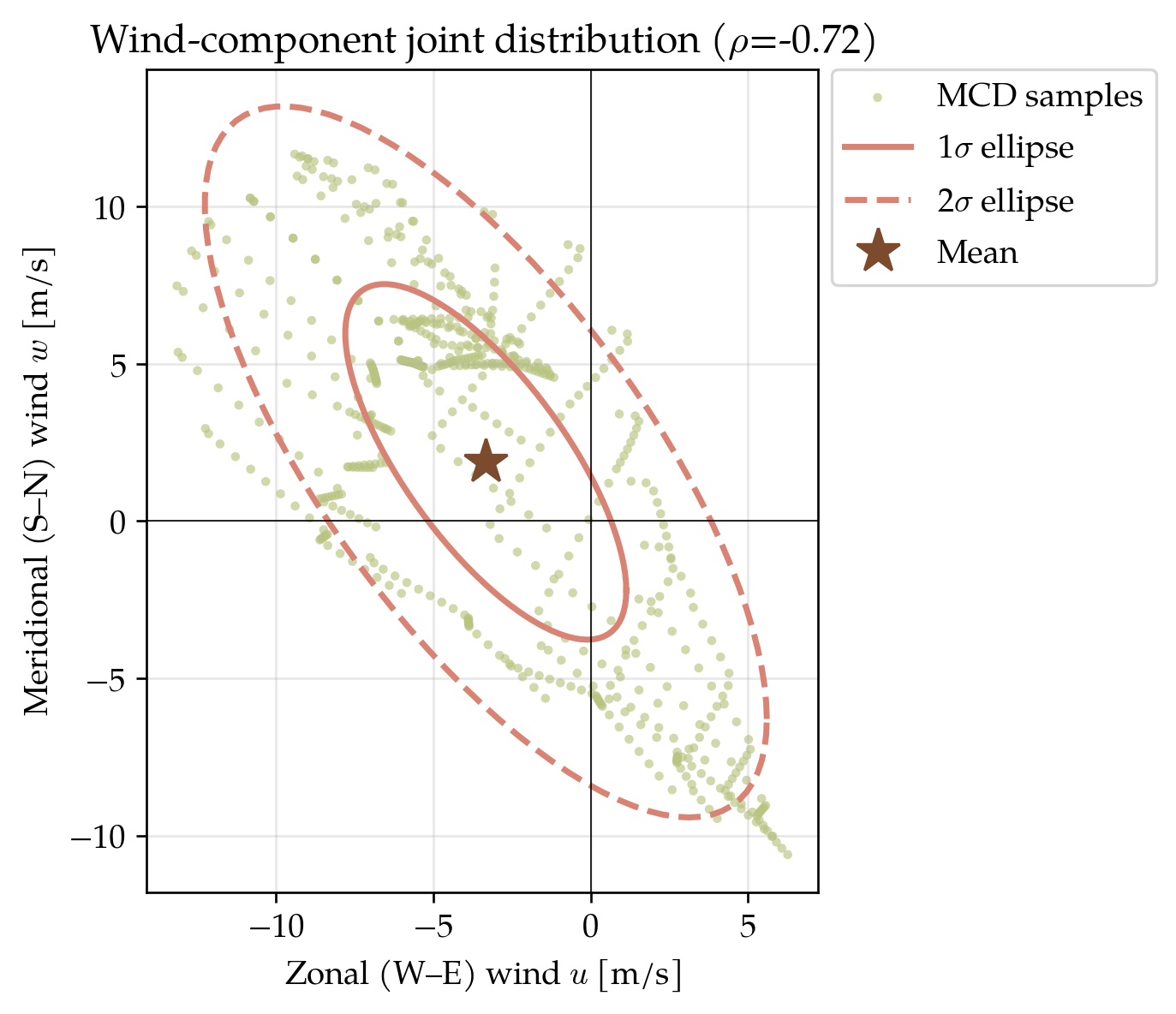}\\(\textbf{b})
\end{minipage}
\end{adjustwidth}
\caption{\hl{Wind} %MDPI: 1. We moved Figure 3 after its first citation, please confirm. 2. Confirm Please change the hyphen (-) into a minus sign (−, “U+2212”) in the figure, e.g., “-1” should be “−1”.Confirm. 
 model from MCD MY34 data at the Curiosity site. (\textbf{a}) Horizontal wind-speed histogram with maximum-likelihood Weibull fit, with~a Kolmogorov--Smirnov statistic KS \mbox{$D = 0.092$}. Because~the 625 states form a structured 25 \hl{$\times$} %MDPI: We revised the letter “x” to a multiplication sign (“×” U+00D7). Please confirm. Confirm
 25 grid in local time and $L_s$ rather than 625 independent samples, we report $D$ as a goodness-of-fit diagnostic and do not quote an independent and identically distributed (iid) significance level; a block bootstrap over $L_s$ columns yields $95\%$ intervals \mbox{$k \in [2.30, 3.24]$}, $\lambda \in [7.69, 9.22]$, wider than the iid parametric intervals \mbox{$k \in [2.57, 2.90]$}, $\lambda \in [8.27, 8.78]$. (\textbf{b}) Joint distribution of the zonal ($u$) and meridional ($w$) wind components with $1\sigma$ and $2\sigma$ bivariate-normal~ellipses.\label{fig:wind}}
\end{figure}

\begin{figure}[H]
\includegraphics[width=0.8\linewidth]{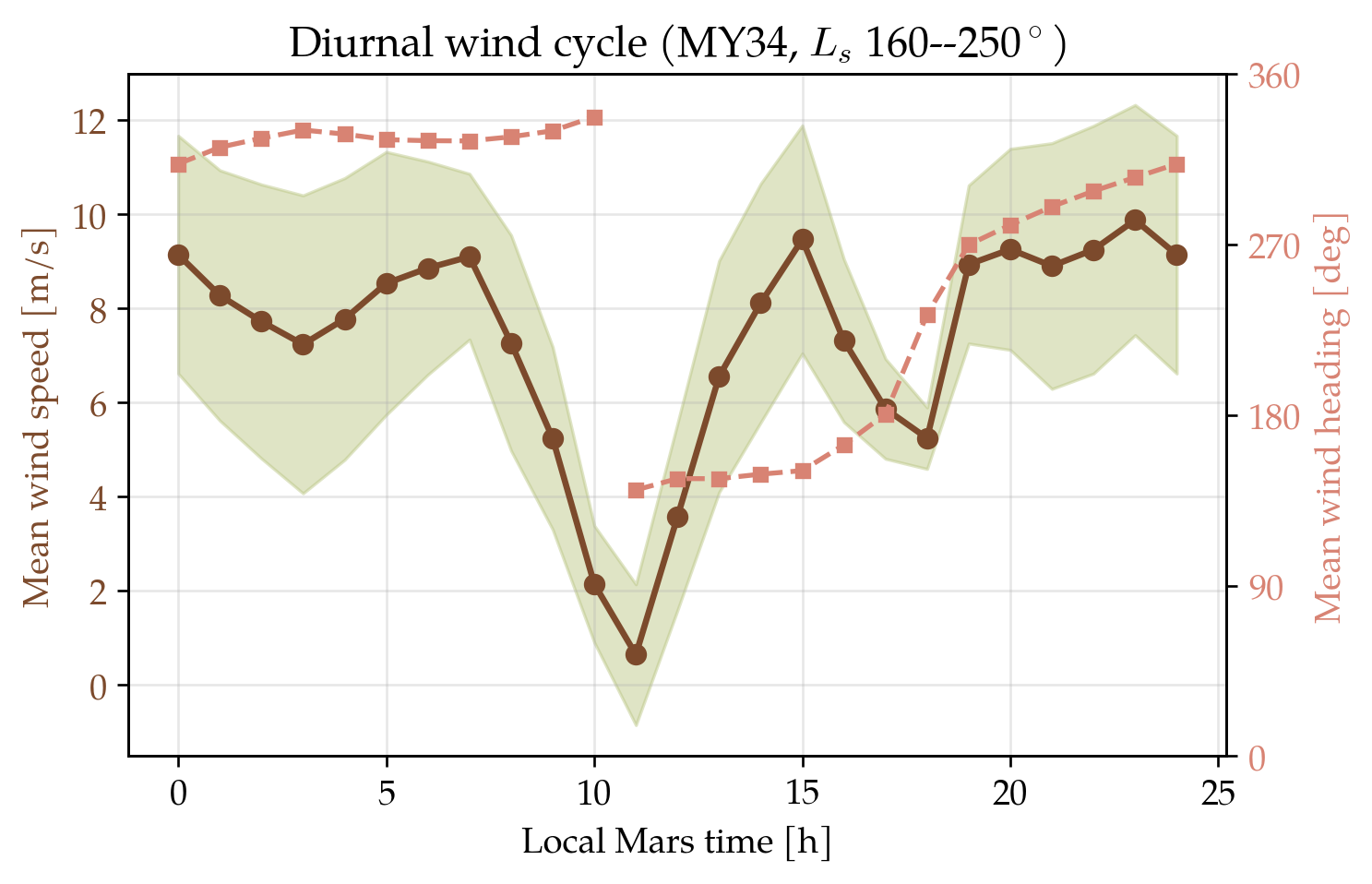}
\caption{\hl{Diurnal} %MDPI: Please change the hyphen (-) between numbers into an en dash (−). Confirm
 cycle of the near-surface wind (MY34, $L_s\,160$--$250^\circ$): mean speed (brown) and mean heading (pink) versus local Mars time. The~shaded band (green) is $\pm1$ standard deviation of speed. The~heading line is broken across the pre-noon calm (hour 11, mean speed 0.65~$\text{m}\,\text{s}^{-1}$), where the mean wind is near zero and its direction is poorly defined (i.e., the heading swings by $\sim$180$^{\circ}$). \label{fig:diurnal}}
\end{figure}

The fitted wind-speed distribution and the bivariate component model are shown in Figure~\ref{fig:wind}. The~Weibull fit ($k=2.72$, $\lambda=8.52~\mathrm{m\,s^{-1}}$) reproduces the speed marginal. The~$(u,w)$ components are strongly anti-correlated, with~a Pearson coefficient $\rho_p=-0.72$ and covariance $\boldsymbol{\Sigma}=\left[\begin{smallmatrix}19.9 & -18.2\\ -18.2 & 31.9\end{smallmatrix}\right]~\mathrm{m^2\,s^{-2}}$. This dependence confirms that an independent speed--direction model would be inappropriate. Removing the hourly mean leaves a gust residual with the much smaller covariance $\boldsymbol{\Sigma}_g=\left[\begin{smallmatrix}2.7 & -2.2\\ -2.2 & 4.5\end{smallmatrix}\right]~\mathrm{m^2\,s^{-2}}$ (standard deviations $1.6$ and $2.1~\mathrm{m\,s^{-1}}$), against~a diurnal mean-speed swing from $0.65$ to $9.9~\mathrm{m\,s^{-1}}$ (Figure~\ref{fig:diurnal}). The~wind field is thus dominated by a coherent, swarm-common diurnal cycle, with~comparatively small rover-to-rover~gusts.

$\boldsymbol{\Sigma}_g$ is the residual of hourly states about the diurnal mean and is dominated by seasonal and storm-phase variation. We therefore treat $\boldsymbol{\Sigma}_g$ as a proxy for the unresolved variability in this model, not as a measured turbulent gust variance. Its amplitude is swept as a sensitivity parameter (Section \ref{sec:conv}). 

The gust residual $\xi$ departs from Gaussian (Shapiro \textit{p} $\leq 10^{-16}$; excess kurtosis $\approx -1$), so the bivariate-normal gust is a tractability assumption for the OU model rather than a fitted distribution. The~bivariate-normal component model, not the Weibull speed marginal, is used in the dynamics; the two are not identical (a general anisotropic correlated Gaussian does not imply a Weibull speed). 
\vspace{-5pt}

\subsection{Mobility~Envelope}

\begin{figure}[H]

\begin{adjustwidth}{-\extralength}{0cm}
\centering %% If there is a figure in wide page, please release command \centering, for Table, ``\textwidth" should be ``\fulllength"
\begin{minipage}{0.65\textwidth}\centering
\includegraphics[width=\linewidth]{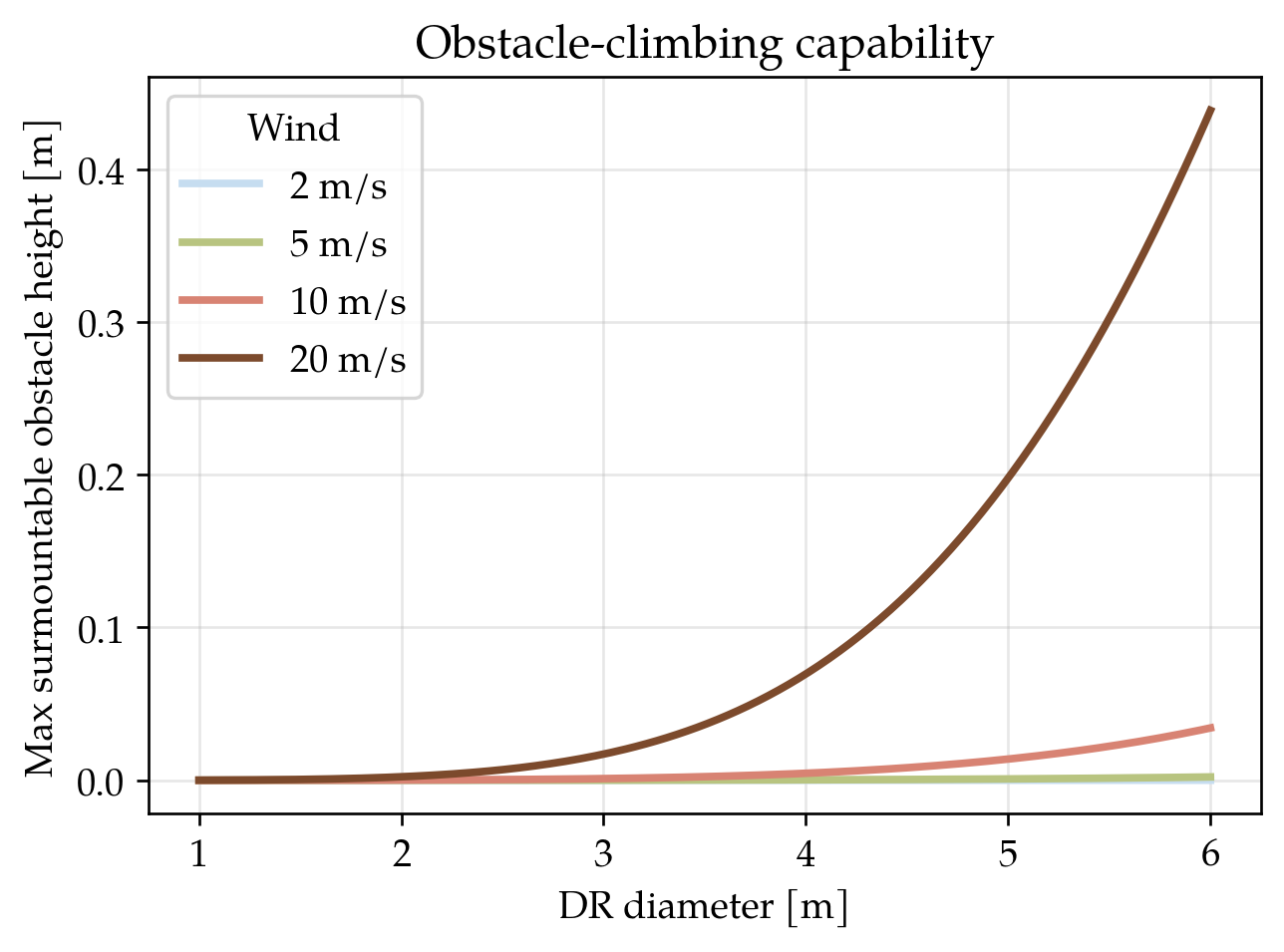}\\(\textbf{a})
\end{minipage}
\begin{minipage}{0.65\textwidth}\centering
\includegraphics[width=\linewidth]{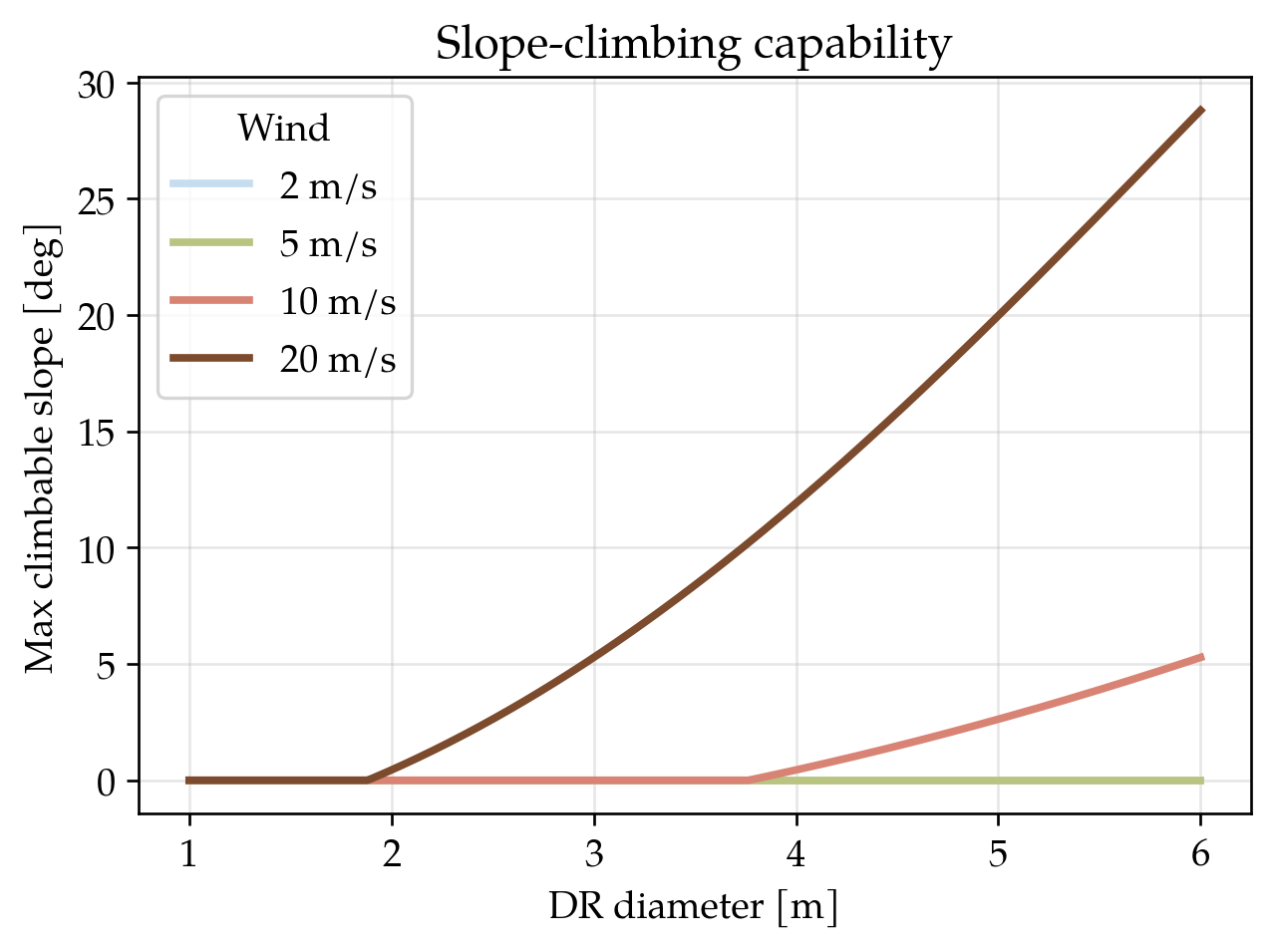}\\(\textbf{b})
\end{minipage}
\end{adjustwidth}
\caption{\hl{DR} %MDPI: There is no blue lines in sub-figure b, please confirm if the labels are correct. Confirm
 mobility versus diameter for winds of 2, 5, 10, and 20~m\,s$^{-1}$. (\textbf{a}) Maximum surmountable obstacle height. (\textbf{b}) Maximum climbable~slope.\label{fig:mobility}}
\end{figure}

Figure~\ref{fig:mobility} evaluates Equations~\eqref{eq:obstacle} and~\eqref{eq:slope} across a spread of DR diameters (1--6~m) for representative winds. The~baseline 4~m DR clears centimetre-scale obstacles---around $1~\mathrm{cm}$ in typical winds and $\sim$7~cm in storm-strength ($20~\mathrm{m\,s^{-1}}$) gusts---and climbs slopes of order $10^\circ$ under strong winds; a larger 6~m rover clears $\sim$0.4~m and $\sim$30$^\circ$ at $20~\mathrm{m\,s^{-1}}$. Since both capabilities scale with $v_{\mathrm{w}}^2$, storm conditions significantly enhance mobility. The~diameter is also a direct mobility--stowage trade-off, with~larger rovers clearing more significant obstacles. We also acknowledge that large-diameter mobility is optimistic, as~areal density and inflation-structure mass are not~scaled.

\subsection{Trajectory and Dispersion~Results}

\begin{figure}[H]

\begin{adjustwidth}{-\extralength}{0cm}
\centering %% If there is a figure in wide page, please release command \centering, for Table, ``\textwidth" should be ``\fulllength"
\includegraphics[width=0.82\fulllength]{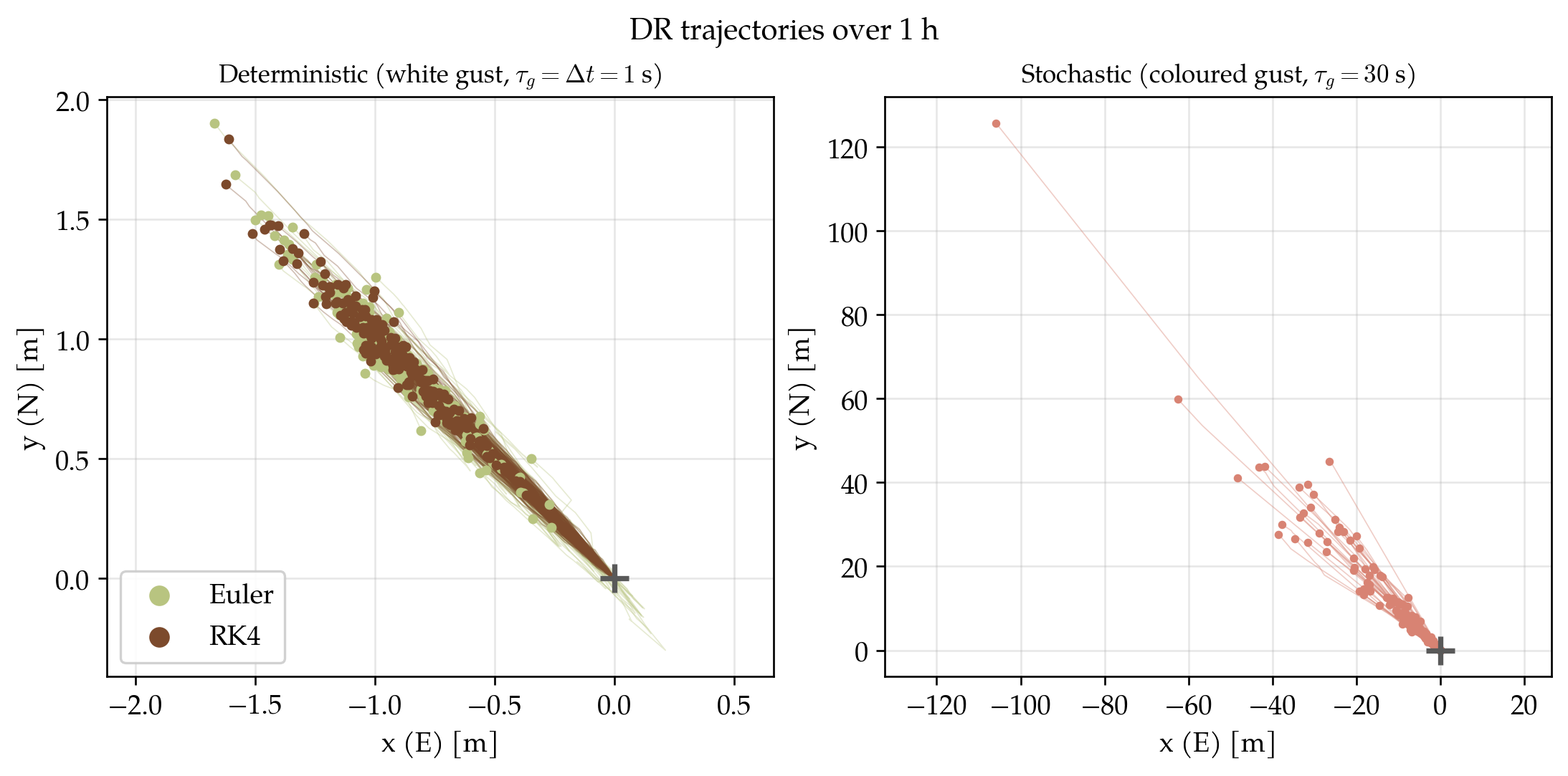}
\end{adjustwidth}
\caption{\hl{DR} %MDPI: Please confirm whether an explanation of the lines and pink dots needs to be added to the figure caption, the same to Figure 7. No need to add.
 trajectory ensembles over one hour under the deterministic white-gust case for both integrators ($\tau_g=\Delta t = 1$~s, \textbf{left}) and a coloured gust ($\tau_g=30$~s, \textbf{right}).\label{fig:trajhour}}
\end{figure}

Hour-long ensembles (\hl{n} %MDPI: Please confirm whether “n” is a variable and should therefore appear in italics. The same below, please check the whole text. n is a variable, confirm
 = 10,000) under mean storm-period wind yield the two-rover separation statistics in Table~\ref{tab:sep} and produce the trajectories shown in Figure~\ref{fig:trajhour}. The~deterministic integrators, which resample the gust as white noise each second, produce mean separations of 0.42~m (Euler) and 0.38~m (RK4). The~stochastic separation depends on the gust correlation time $\tau_g$ and is quantified in Table~\ref{tab:tau}.

\begin{figure}[H]

\begin{adjustwidth}{-\extralength}{0cm}
\centering %% If there is a figure in wide page, please release command \centering, for Table, ``\textwidth" should be ``\fulllength"
\includegraphics[width=0.82\fulllength]{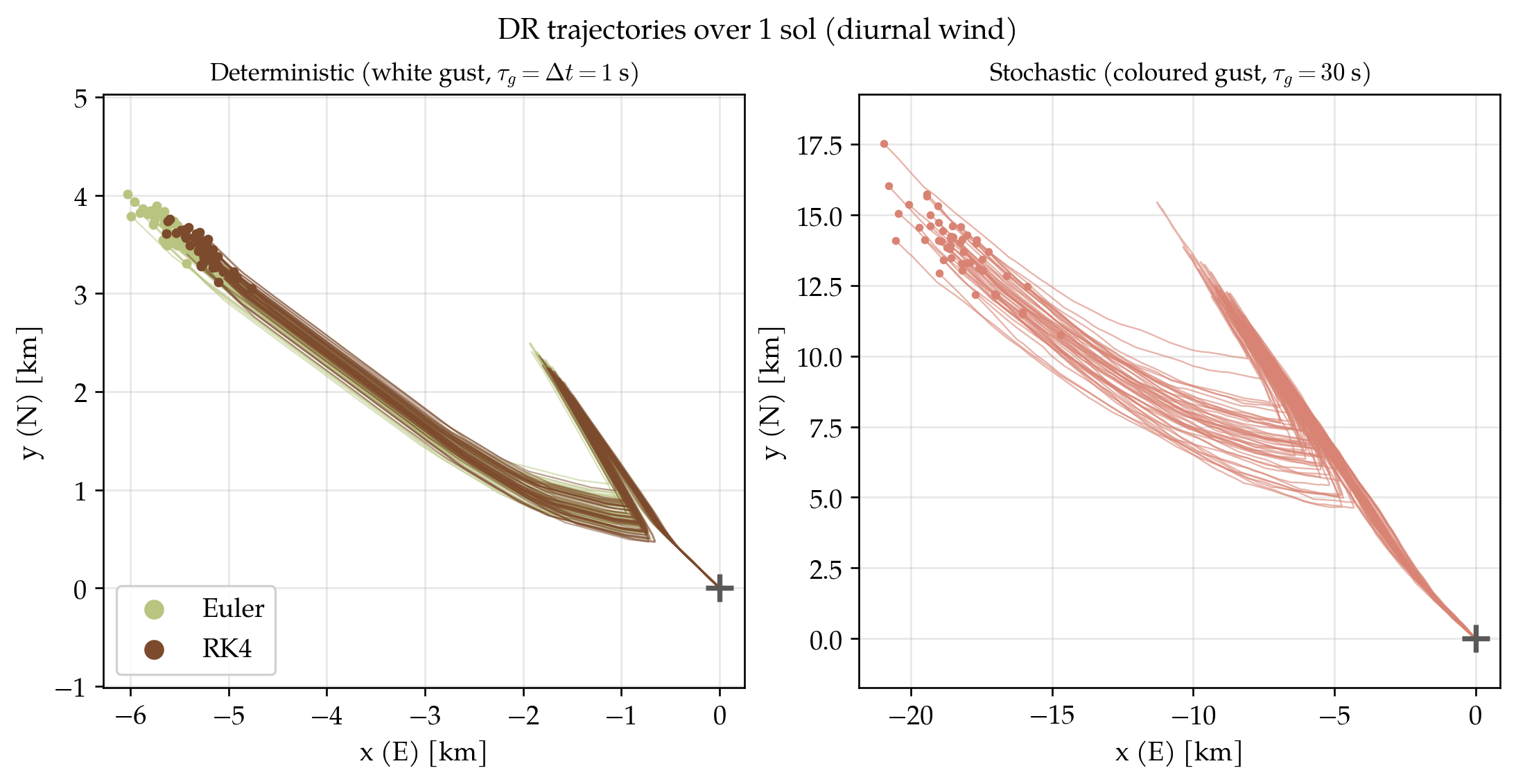}
\end{adjustwidth}
\caption{\hl{DR} %MDPI: We moved Figure 7 after its first citation, please confirm. Confirm
 trajectories over one sol under the deterministic white-gust case for both integrators ($\tau_g=\Delta t = 1$~s, \textbf{left}) and a coloured gust ($\tau_g=30$~s, \textbf{right}), showing the diurnal~reversal.\label{fig:trajsol}}
\end{figure}

Over a full sol with the diurnal wind cycle (Figure~\ref{fig:trajsol}), the~deterministic drift carries the swarm $\sim$11~km with a characteristic looping path as the day- and night-time regimes reverse. The~two-rover separation shows similar patterns as over an hour: 0.44~km on average (99th percentile 1.3~km) under the deterministic, white-gust case, rising to $1.4$--$3.3$~km on average ($99$th percentile $4.1$--$10.1$~km) for stochastic, gust correlation times of $10$--$60$~s (Table~\ref{tab:tau}).

\subsection{Communication-Network~Connectivity}\label{sec:netresults}
\begin{figure}[H]

\begin{adjustwidth}{-\extralength}{0cm}
\centering %% If there is a figure in wide page, please release command \centering, for Table, ``\textwidth" should be ``\fulllength"
\includegraphics[width=0.85\fulllength]{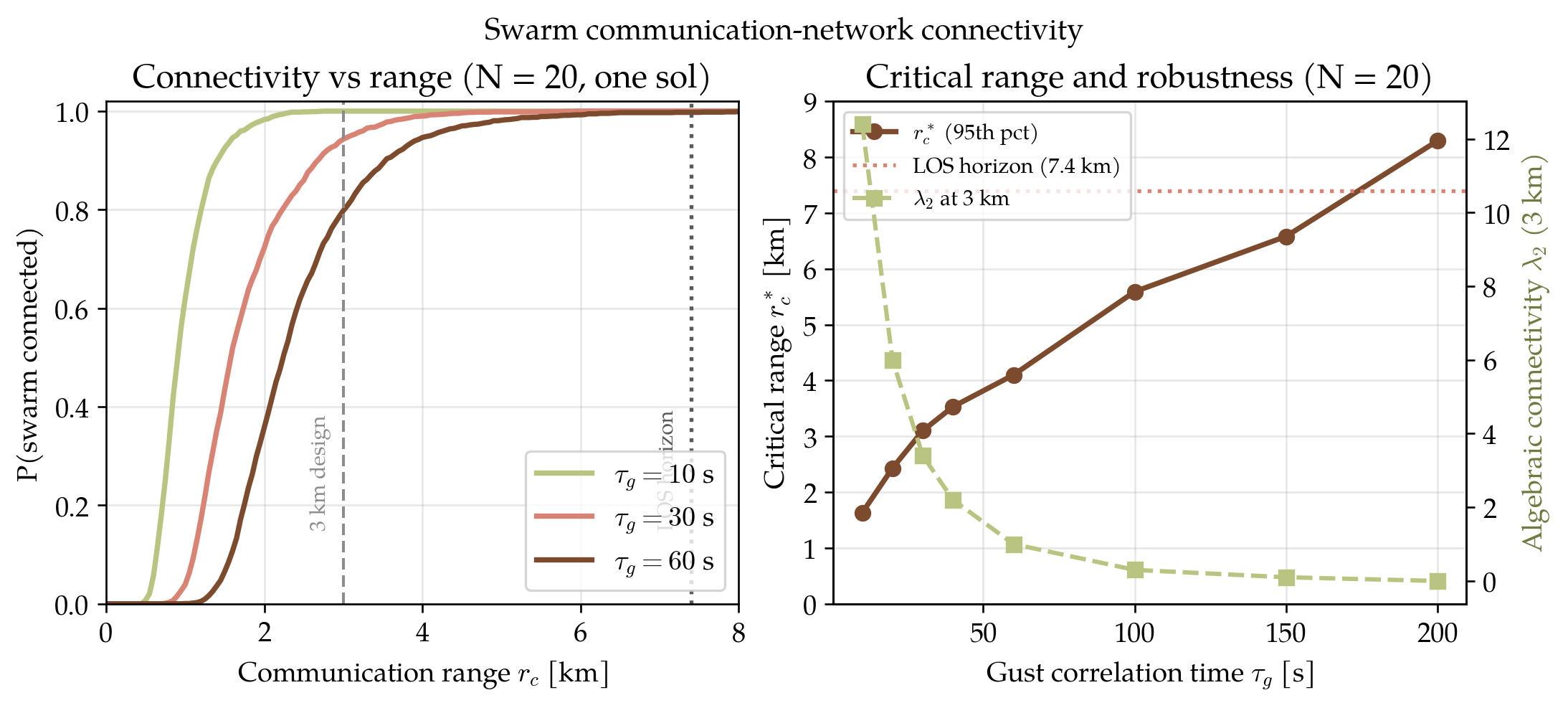}
\end{adjustwidth}
\caption{\hl{Swarm} %MDPI: 1. We moved Figure 8 after its first citation, please confirm. 2. Please confirm whether the overlapping content in this figure affects scientific understanding and if it does, please revise it. Confirm. No need to revise.
 communication-network connectivity ($N=20$, end of one sol). (\textbf{a}) Probability the swarm forms a single connected component versus per-rover communication range $r_c$, for~gust correlation times $\tau_g=10,30,60$~s; the dotted line marks the $7.4$~km LOS horizon. (\textbf{b}) Critical range $r_c^\star$ (95th percentile) versus $\tau_g$, against~the LOS~horizon.\label{fig:netconn}}
\end{figure}

Beyond the pairwise separation, we evaluate the connectivity of the swarm modelled as
a communication graph (Section~\ref{sec:trajmethods}), at~the most-dispersed end-of-sol
state. Figure~\ref{fig:netconn}a presents the probability that a $20$-rover swarm forms a
single connected component as a function of the per-rover communication range $r_c$; the
transition is sharp, and~the critical range $r_c^\star$ (the $95$th percentile of the
longest minimum-spanning-tree edge) rises from $1.6$~km at $\tau_g=10$~s to $4.1$~km at
$\tau_g=60$~s (Figure~\ref{fig:netconn}b, Table~\ref{tab:net}). In~every case, we find  $r_c^\star$ to be below the $7.4$~km LOS horizon for one sol. Evaluated through time rather than only at the end of the sol, the~swarm is connected at the 7.4~km horizon for essentially the whole sol at these correlation times, and~at the 3~km design range for a fraction 1.00, 1.00, 0.99, 0.98, and 0.93 of the sol for $\tau_g = 10, 20, 30, 40$, and 60~s, respectively. At~the $3$~km design range and $\tau_g=60$~s (the least-connected plausible case), $42\%$ of swarms remain connected for the entire sol; when a disconnection occurs, the~longest single outage is $\sim$1~h on average ($95$th percentile $\sim$6~h), which sets the onboard data-buffering requirement. At~the $7.4$~km LOS horizon outages are~negligible.

\vspace{-6pt}
\begin{table}[H]
\caption{\hl{Swarm} %MDPI: We moved Table 5 after its first citation, please confirm. Confirm.
 communication-network metrics versus gust correlation time $\tau_g$ ($N=20$, static-threshold baseline, independent gusts $\ell=0$). $r_c^\star$ is the $95$th percentile of the longest minimum-spanning-tree edge (end of sol); $\lambda_2$ is the algebraic connectivity evaluated at the $3$~km design range (at the LOS range it saturates to the complete-graph value and is uninformative); $P_{\text{conn}}$ is the probability the swarm is a single connected component at $7.4$~km and $3$~km; ``frac.\ sol'' is the mean fraction of the sol connected at $3$~km. $T^\star=(7.4/r_c^\star)^2$ is a heuristic diffusive extrapolation.
\label{tab:net}}
\small
\begin{tabularx}{\textwidth}{c CCccc c}
\toprule
\textbf{$\bm{\tau_g}$ [s]} & \textbf{$\bm{r_c^\star}$ [km]} & \textbf{$\bm{\lambda_2}$ (3 km)} & \textbf{$\bm{P_{\text{conn}}}$ (7.4 km)} & \textbf{$\bm{P_{\text{conn}}}$ (3 km)} & \textbf{frac.\ sol (3 km)} & \textbf{$\bm{T^\star}$ [sols]}\\
\midrule
10  & 1.6 & 12.4 & 1.00 & 1.00 & 1.00 & 21\\
20  & 2.4 & 6.0  & 1.00 & 0.99 & 1.00 & 9\\
30  & 3.1 & 3.4  & 1.00 & 0.94 & 0.99 & 6\\
40  & 3.5 & 2.2  & 1.00 & 0.90 & 0.98 & 4\\
60  & 4.1 & 1.0  & 1.00 & 0.80 & 0.93 & 3\\
100 & 5.6 & 0.3  & 0.99 & 0.53 & 0.84 & 2\\
150 & 6.6 & 0.1  & 0.97 & 0.32 & 0.72 & 1\\
200 & 8.3 & 0.0  & 0.91 & 0.11 & 0.61 & <$1$\\
\bottomrule
\end{tabularx}
\end{table}

Reported at the 3~km design range, the~algebraic connectivity falls from $\lambda_2 = 12.4$ at $\tau_g = 10$~s to 1.0 at $\tau_g = 60$~s, meaning the network thins from a dense mesh towards a sparse but connected graph as $\tau_g$ grows. The~critical range decreases with swarm size, from~4.4~km at N = 5 to 2.8~km at N = 50 (for $\tau_g=30$~s). Extending the \hl{sweep,} %MDPI: %MDPI: Please carefully check variable formatting (italic, bold, subscript, uppercase, etc.) throughout the manuscript to ensure the formatting is consistent and revise if needed.
 $r_c\star$ reaches the 7.4~km horizon near $\tau_g \approx 180-200$~s, bracketing the loss of one-sol connectivity. {Because}~$r_c\star$ grows diffusively, a~heuristic extrapolation of $r_c\star \propto \sqrt{t}$ provides a connectivity horizon T of $\sim$21 sols at $\tau_g = 10$~s, falling to $\sim$3 sols at $\tau_g = 60$~s. 

\begin{figure}[H]

\begin{adjustwidth}{-\extralength}{0cm}
\centering %% If there is a figure in wide page, please release command \centering, for Table, ``\textwidth" should be ``\fulllength"
\includegraphics[width=0.9\fulllength]{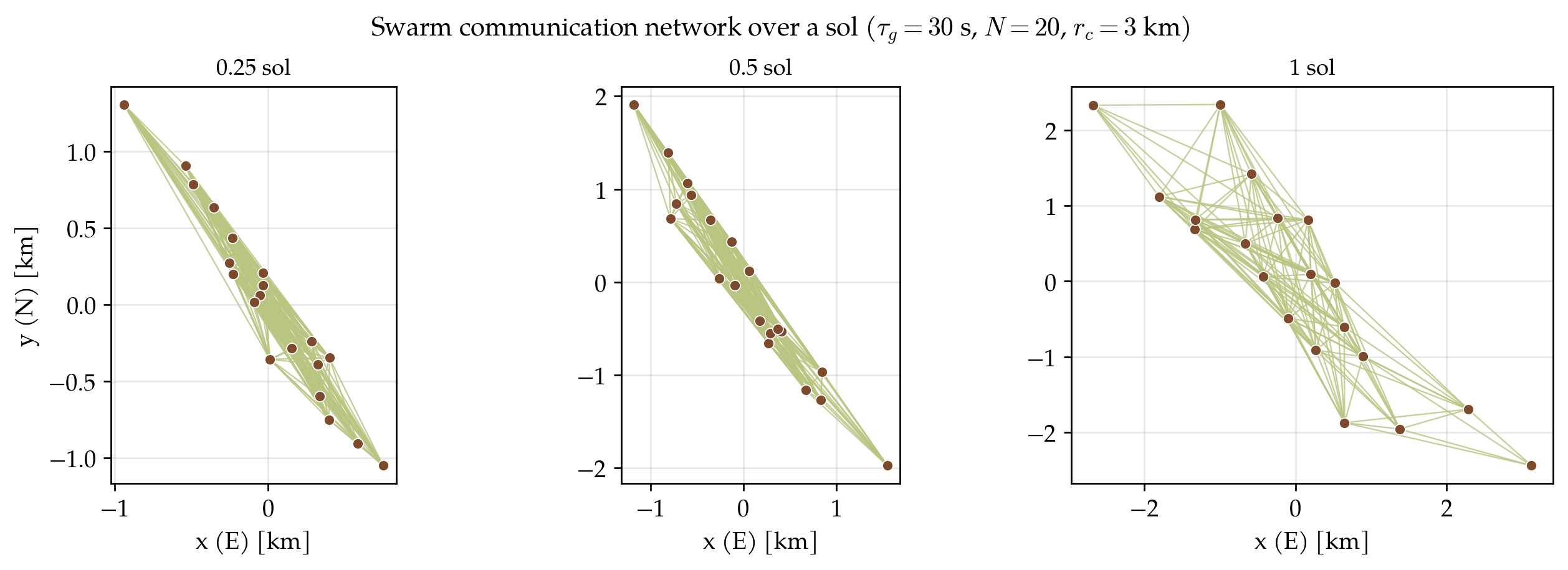}
\end{adjustwidth}
\caption{\hl{Communication} %MDPI: 1. Please cite this figure in the text and ensure that the first citation of each figure appears in numerical order. 2. Please confirm whether an explanation of the lines and dots needs to be added to the figure caption. No need to add, figure is cited.
 graph of one $20$-rover swarm at $0.25$, $0.5$, and $1$ sol ($\tau_g=30$~s, range $r_c=3$~km), centred on the swarm centroid. Edges are viable single hops; the relay network stretches but stays connected as the swarm disperses. A~3~km design range limit is marked by a dashed line in the left panel, and~$\lambda_2$ is overlayed at 3~km in the right~panel. \label{fig:netsnap}}
\end{figure}

\begin{figure}[H]

\begin{adjustwidth}{-\extralength}{0cm}
\centering %% If there is a figure in wide page, please release command \centering, for Table, ``\textwidth" should be ``\fulllength"
\includegraphics[width=0.83\fulllength]{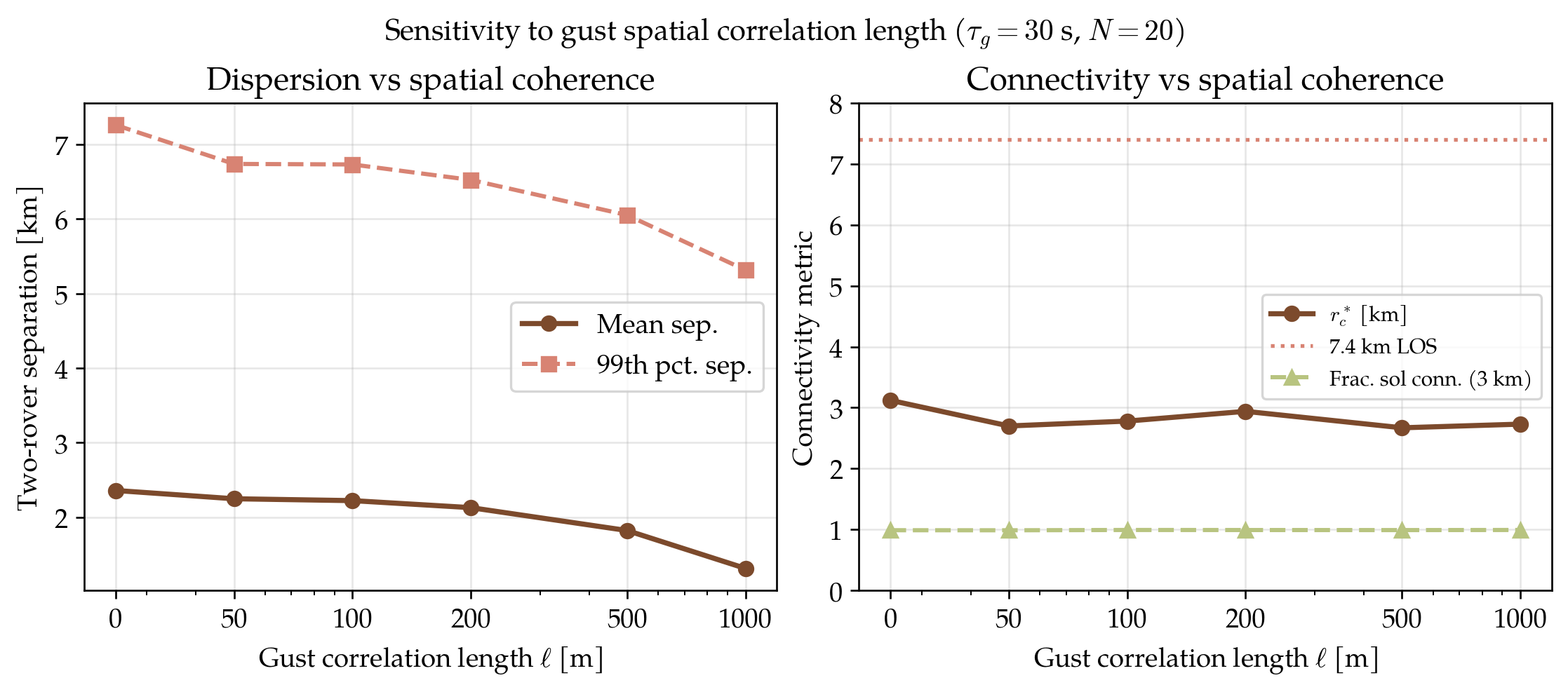}
\end{adjustwidth}
\caption{Sensitivity of dispersion and connectivity to the gust spatial correlation length $\ell$ ($\tau_g=30$~s, $N=20$). The~gust field is assigned an exponential spatial coherence, $C(\delta)=\exp(-\delta/\ell)$, so that two rovers separated by $\delta$ share a fraction of their forcing; $\ell=0$ is the spatially independent case, retained as the conservative bound. \textbf{Left}: Mean and 99th-percentile two-rover separation [km] versus \textls[-10]{$\ell$ (logarithmic axis). \textbf{Right}: Critical communication range $r_c^\star$ [km], the~$7.4$~km LOS horizon (dotted), and}}   \label{spatial_sensitivity}
\end{figure}

The spatial coherence of the gust ($C(\delta) =\exp(-\delta/\ell)$; Figure~\ref{spatial_sensitivity} and Table~\ref{tab:spatial}) reduce the two-rover mean separation monotonically from 2.36~km ($\ell=0$, independent) to 1.31~km ($\ell=1000$~m) at $\tau_g = 30$~s and never reduce connectivity ($P_\mathrm{conn}$ at 7.4~km stays 1.00). The~feasibility result is therefore robust to plausible spatial correlation: the mean separation falls, while the critical range shows no systematic trend and the connected fraction is unchanged (Table~\ref{tab:spatial}).
\vspace{-6pt}
{\captionof*{figure}{ the fraction of the sol connected at the $3$~km design range, versus $\ell$. Greater spatial coherence reduces relative dispersion and does not degrade connectivity (the critical range shows no systematic trend and the connected fraction stays at 0.99), so the independent-gust result is~conservative.}}
\vspace{6pt}

\begin{table}[H]
\caption{Sensitivity of one-sol dispersion and connectivity to the gust spatial
correlation length $\ell$ (exponential coherence $C(\delta)=\exp(-\delta/\ell)$;
$\tau_g=30$~s, $N=20$). $\ell=0$ is the independent-gust conservative bound. Increasing spatial coherence reduces relative dispersion monotonically and does not degrade connectivity---the critical range shows no systematic trend ($2.67$--$3.12$~km) and the connected fraction stays at $0.99$---so the feasibility conclusion is robust to plausible spatial correlation.\label{tab:spatial}}
\begin{tabularx}{\textwidth}{cCCCC}
\toprule
\textbf{$\bm{\ell}$ [m]} & \textbf{Mean $\bm{D}$ [km]} & \textbf{99th $\bm{D}$ [km]} & \textbf{$\bm{r_c^\star}$ [km]} & \textbf{Frac.\ Sol (3 km)}\\
\midrule
0    & 2.36 & 7.26 & 3.12 & 0.99\\
50   & 2.25 & 6.74 & 2.70 & 0.99\\
100  & 2.22 & 6.73 & 2.78 & 0.99\\
200  & 2.13 & 6.52 & 2.94 & 0.99\\
500  & 1.82 & 6.05 & 2.67 & 0.99\\
1000 & 1.31 & 5.31 & 2.73 & 0.99\\
\bottomrule
\end{tabularx}
\end{table}

\begin{figure}[H]

\begin{adjustwidth}{-\extralength}{0cm}
\centering %% If there is a figure in wide page, please release command \centering, for Table, ``\textwidth" should be ``\fulllength"
\includegraphics[width=0.9\fulllength]{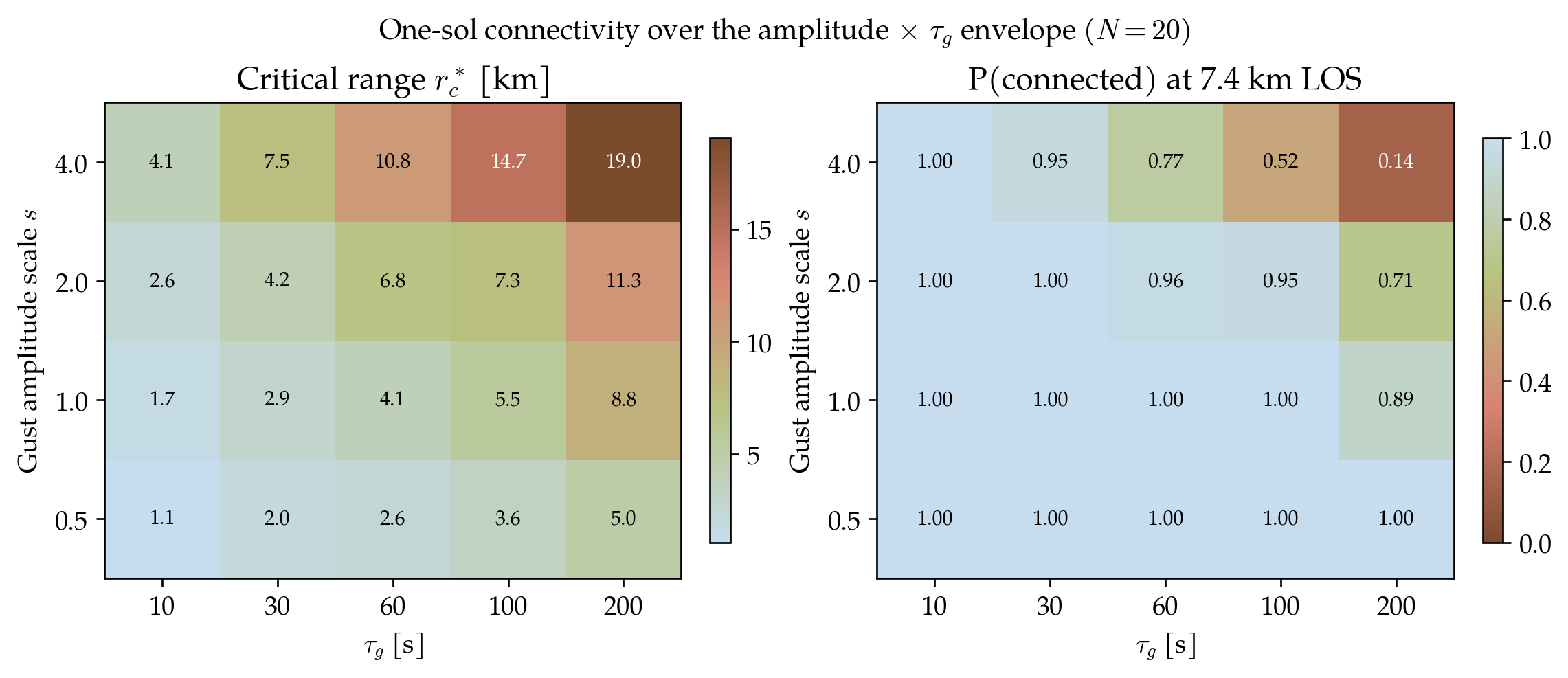}
\end{adjustwidth}
\caption{One-sol swarm connectivity over the gust amplitude--correlation-time envelope ($N=20$). The~stationary gust covariance is scaled by a factor $s$ (rows), and the correlation time $\tau_g$ is varied (columns), both treated as bounded unknowns. \textbf{Left}: Critical communication range $r_c^\star$ [km]---the smallest per-rover range that connects the swarm, taken as the 95th percentile of the longest edge of the swarm's Euclidean minimum spanning tree. \textbf{Right}: Probability that the swarm forms a single connected component at the $7.4$~km geometric line-of-sight (LOS) horizon. In~both panels, darker cells denote poorer connectivity (larger $r_c^\star$, lower $P$); the swarm remains connected within the LOS horizon across most of the plausible envelope, failing only where the gust amplitude and correlation time are simultaneously~large.\label{heatmap}}
\end{figure}

Because the gust amplitude $\Sigma_g$ is itself uncertain, we report connectivity over a two-parameter envelope (amplitude scale, $s$, and~correlation time, $\tau_g$; Table~\ref{tab:env} and Figure~\ref{heatmap}). The~one-sol swarm stays connected within the 7.4~km horizon ($r_c\star < 7.4$~km, $P_\mathrm{conn} \ge 0.95$) for all $s \le 2$ with $\tau_g \le 60$~s and along the baseline $s = 1$ up to $\tau_g \approx 150$~s. Connectivity is lost only when the amplitude and correlation time are simultaneously large. Tables~\ref{tab:net} and~\ref{tab:env} are independent Monte-Carlo runs; the $95$th-percentile $r_c^\star$ carries a sampling uncertainty of $\approx$$\pm0.04$~km at $\tau_g=10$~s and $\approx$$\pm0.15$~km at $\tau_g=30$~s ($95\%$ bootstrap CI over the $2000$ swarms), which accounts for the small differences between the tables (e.g.,\ $1.6$ \hl{vs.} %MDPI: We add Abbreviation point for vs, the same below, please confirm. Confirm
 $1.7$~km at $10$~s and $3.1$ vs. $2.9$~km at $30$~s).

\begin{table}[H]
\caption{One-sol connectivity over the two-parameter uncertainty envelope (amplitude scale $s$ on $\Sigma_g$ $\times$ gust correlation time $\tau_g$; $N=20$, $\ell=0$). Each cell provides the critical range $r_c^\star$ [km] and, in~parentheses, the~probability the swarm is connected within the $7.4$~km LOS horizon. The~swarm stays connected ($r_c^\star<7.4$~km, $P_{\text{conn}}\ge0.95$) across the plausible region $s\le2,\ \tau_g\le60$~s and along $s=1$ up to $\tau_g\approx150$~s; connectivity is lost only for simultaneously large amplitude and long correlation~time.\label{tab:env}}
\begin{tabularx}{\textwidth}{cCCCCC}
\toprule
\textbf{$\bm{s\backslash\tau_g}$ [s]} & \textbf{10} & \textbf{30} & \textbf{60} & \textbf{100} & \textbf{200}\\
\midrule
0.5 & 1.1 (1.00) & 2.0 (1.00) & 2.6 (1.00) & 3.6 (1.00) & 5.0 (1.00)\\
1.0 & 1.7 (1.00) & 2.9 (1.00) & 4.1 (1.00) & 5.5 (1.00) & 8.8 (0.89)\\
2.0 & 2.6 (1.00) & 4.2 (1.00) & 6.8 (0.97) & 7.3 (0.95) & 11.3 (0.71)\\
4.0 & 4.1 (1.00) & 7.5 (0.95) & 10.8 (0.77) & 14.7 (0.52) & 19.0 (0.14)\\
\bottomrule
\end{tabularx}
\end{table}

%=================================================================
\section{Discussion}\label{sec:discussion}

\subsection{Numerical Convergence and Sensitivity~Analysis}
\label{sec:conv}

\begin{figure}[H]
\includegraphics[width=0.77\linewidth]{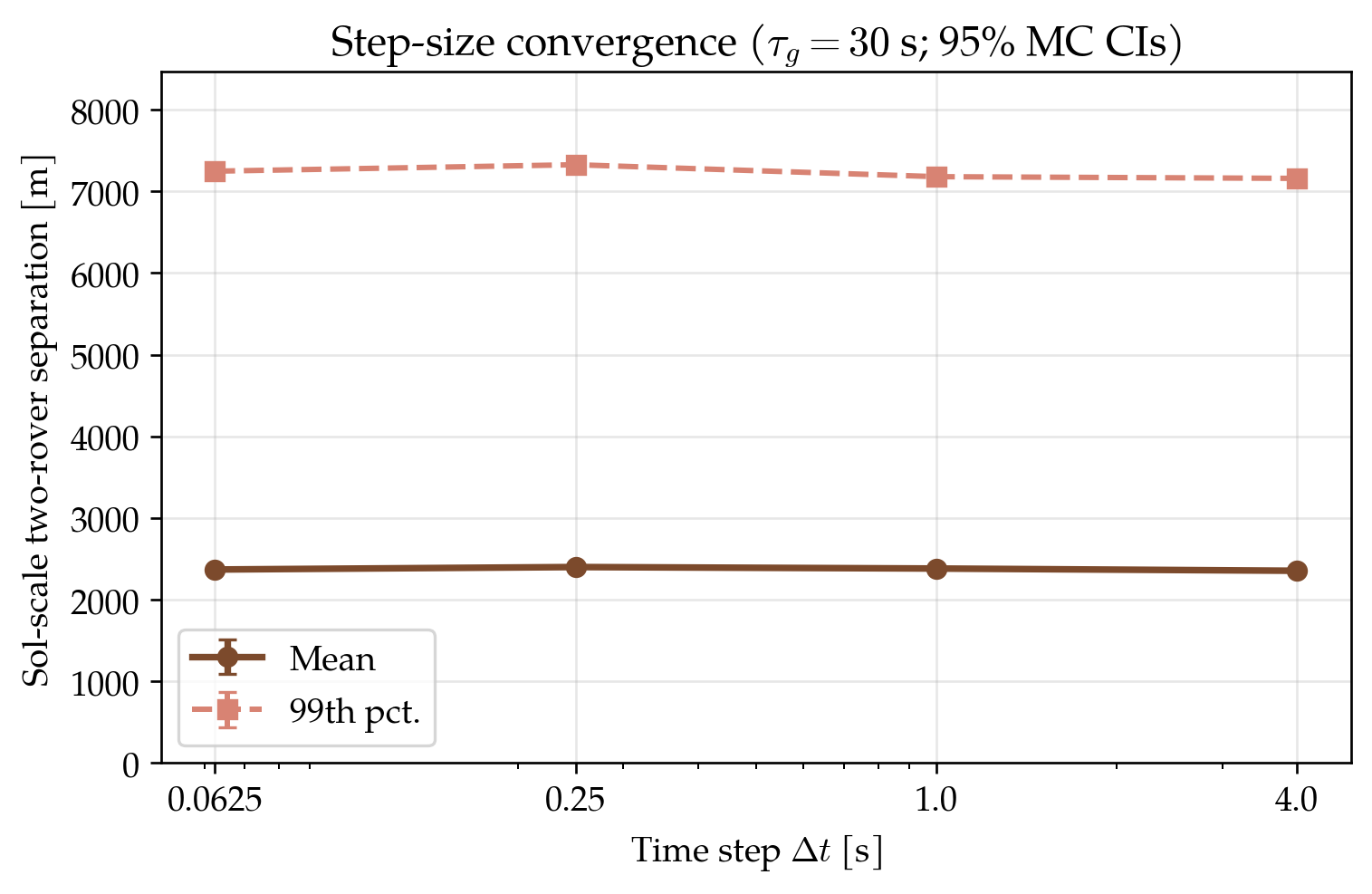}
\caption{\hl{Step-size} %MDPI: We moved Figure 12 after its first citation, please confirm. Confirm
 convergence of the sol-scale separation (coloured gust, $\tau_g=30$~s).\label{fig:conv}}
\end{figure}

The three integrators play complementary roles. The~deterministic Euler and RK4 schemes resample the gust as white noise, so successive gust bursts cancel, and separation stays small (below $1$~m over an hour). The~stochastic scheme integrates the coloured gust, whose correlation time $\tau_g$ is left unconstrained. We verified convergence in the time step by repeating the sol-scale simulation ($\tau_g=30$~s) at $\Delta t = 4, 1, 0.25$, and~$0.0625$~s with matched random seeds and 95\% Monte-Carlo confidence intervals (Table~\ref{tab:conv} and Figure~\ref{fig:conv}). Across this $64\times$ refinement, the~sol-scale mean separation varies by less than $2\%$ (Table~\ref{tab:conv}). This variation across $\Delta t$ is far smaller than the variation across $\tau_g$ (Table~\ref{tab:tau}, a~factor of several over $10$--$300$~s): the dispersion is limited by the physical, unconstrained gust correlation time rather than by numerical resolution, which the hourly, single-point MCD data cannot resolve (Section~\ref{gausvsweib}).
\vspace{-4pt}
\vspace{-6pt}
\begin{table}[H]
\caption{Step-size convergence of the sol-scale two-rover separation ($\tau_g=30$~s), matched seeds, with~$95\%$ Monte-Carlo confidence intervals. Across a $64\times$ refinement the mean varies by ${<}2\%$ and the $99$th percentile by ${\sim}2\%$, comparable to the sampling CIs, so $\Delta t=1$~s is~adequate.\label{tab:conv}}
\begin{tabularx}{\textwidth}{cCCC}
\toprule
\textbf{$\bm{\Delta t}$ [s]} & \textbf{Steps} & \textbf{Sol Mean [m] (95\% CI)} & \textbf{Sol 99th [m] (95\% CI)}\\
\midrule
4.00   & 22{,}200    & 2354 (2348--2359) & 7160 (7123--7193)\\
1.00   & 88{,}800    & 2381 (2376--2387) & 7180 (7151--7212)\\
0.25   & 355{,}200   & 2399 (2394--2405) & 7327 (7290--7363)\\
0.0625 & 1{,}420{,}800 & 2369 (2363--2375) & 7248 (7197--7291)\\
\bottomrule
\end{tabularx}
\end{table}

\begin{figure}[H]

\begin{adjustwidth}{-\extralength}{0cm}
\centering %% If there is a figure in wide page, please release command \centering, for Table, ``\textwidth" should be ``\fulllength"
\includegraphics[width=0.85\fulllength]{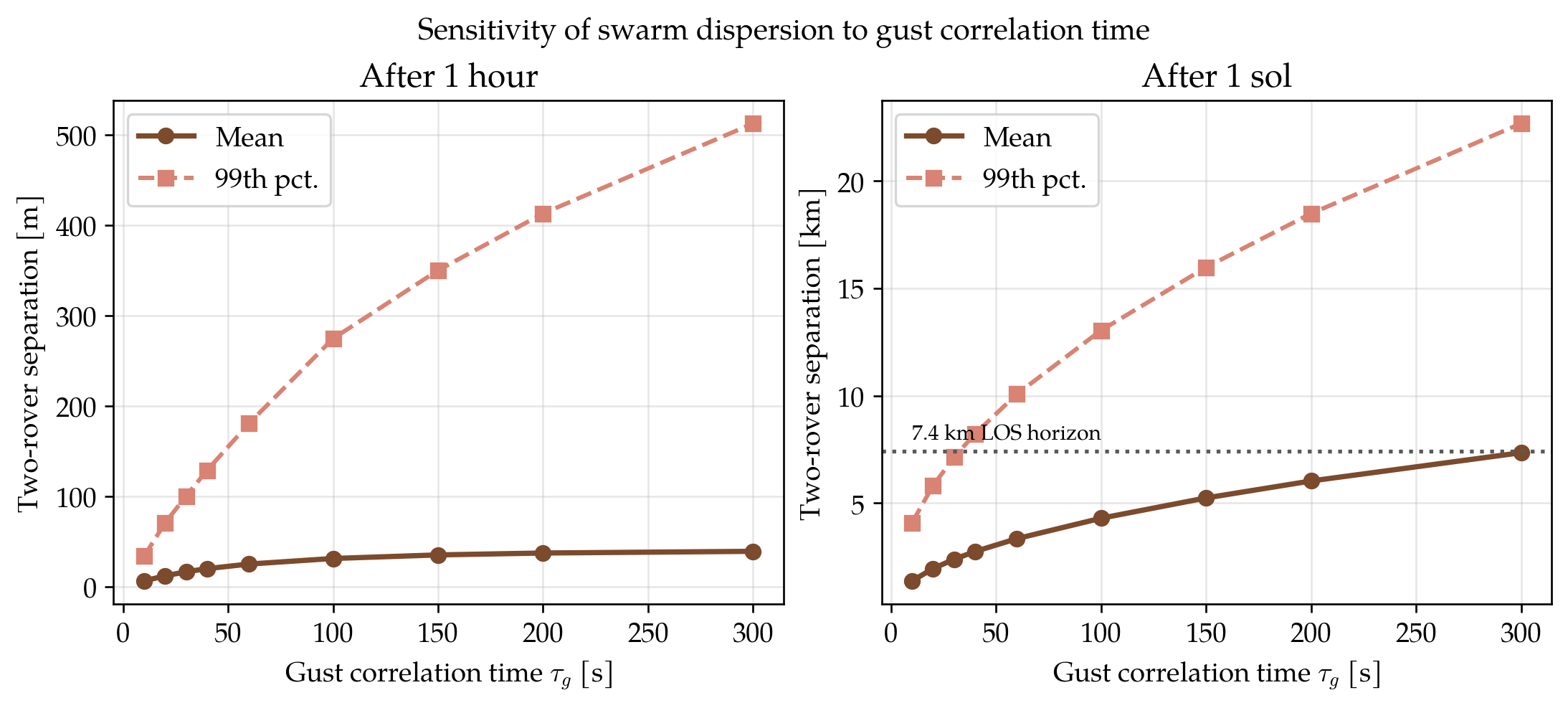}
\end{adjustwidth}
\caption{\hl{Sensitivity} %MDPI: 1. We moved Figure 13 after its first citation, please confirm. 2. Please confirm whether the overlapping content in this figure affects scientific understanding and if it does, please revise it. Confirm, no revision needed.
 of two-rover separation to the gust correlation time $\tau_g$, after~one hour (\textbf{left}) and one sol (\textbf{right}; dotted line is the $7.4$~km LOS horizon). Solid: mean; dashed: $99$th~percentile.\label{fig:tau}}
\end{figure}

The gust correlation time $\tau_g$ is unconstrained by the hourly data yet governs how far neighbours drift apart. Sweeping it over $10$--$300$~s
(Table~\ref{tab:tau} and Figure~\ref{fig:tau}), the~sol-scale two-rover separation grows approximately as $\sqrt{\tau_g}$ (from $1.4$ to $3.3$~km over $\tau_g=10$--$60$~s, a~ratio of $2.4$ against $\sqrt{6}\approx2.45$); the hour-scale mean grows faster, from~$6$ to $25$~m (a ratio of $\sim$4.2), because, at~the hour scale, the~motion is still threshold-gated and has not reached the diffusive regime. The~white-gust case ($\tau_g=\Delta t=1$~s) gives $\sim$0.4~m and $0.44$~km. The~sol-scale $99$th-percentile separation stays within the $7.4$~km LOS horizon for $\tau_g\lesssim30$\hl{--}%MDPI: We changed this to an en dash, the same below, please confirm. Confirm
35~s, and the sol mean up to $\tau_g \approx 300$~s; at the swarm level, $r_c\star$ reaches the horizon near $\tau_g\approx 180$\hl{--}200~s. The~physically plausible band is roughly 10\hl{--}100~s because the rover drag-relaxation time is $m/(\rho C_\mathrm{d}AU) \approx 21$~s at the rolling-threshold wind speed, and~Martian daytime boundary-layer eddy turnover is tens of seconds to a few minutes. The~100--300~s is retained as an upper-bounding stress~test.\vspace{-3pt}

\subsection{Gaussian vs. Weibull Forcing and the Limits of the OU~Model}
\label{gausvsweib}
Adopting a bivariate normal distribution for the wind components, rather than an independent or copula-coupled Weibull for $(v,\phi)$, is what makes the OU formulation tractable: a classical OU process presumes Gaussian (Brownian) increments, so a Weibull-based forcing would require either transforming the variables into an approximately Gaussian space or generalising Equation~\eqref{eq:sde} to non-Gaussian, L\'evy-driven noise~\citep{barndorff2002}. Both routes complicate the drift and diffusion and raise the computational burden~\citep{maller2009}, and~a skewed marginal would additionally displace the mean-reversion target from the Gaussian centre. Embedding a realistic speed--direction dependence in that setting would further demand a copula~\citep{krupskii2013}, whereas in the Gaussian component model the dependence is carried, exactly and for free, by~the off-diagonal terms of $\boldsymbol{\Sigma}$. The~component representation also sidesteps the bimodality of the wrapped heading, which no single Gaussian on $(v,\phi)$ can~capture. 

The simplification that remains lies within the calibration of the model. The~gust is integrated as a coloured, mean-reverting process with a finite correlation time $\tau_g$, and~the exact OU transition keeps its stationary covariance independent of the step size. The~hourly MCD cannot supply a value for $\tau_g$: it constrains the mean field $\bar{\mathbf{V}}_{h(t)}$ and the residual covariance $\Sigma_g$ but not the timescale on which the residual decorrelates. We therefore treat $\tau_g$ as a bounded unknown and sweep it (Section \ref{sec:conv}) rather than fixing it; calibrating it would require sub-hourly in situ wind data, such as the 1 Hz REMS record at Gale before the sensor failure~\citep{viudez2019}. 

\subsection{Communication~Feasibility}\label{sec:comm}
The mission requires only that each DR remain within radio range of its nearest neighbour, so that data can be relayed hop-by-hop across the swarm and onward to a lander or orbiter. In~the white-gust case ($\tau_g=\Delta t = 1$~s), the~sol-scale two-rover separation is negligible for this purpose ($99$th percentile $\sim$1.3~km; Table~\ref{tab:tau}). For~the plausible correlation times $\tau_g=10$--$60$~s, it grows to $1.4$--$3.3$~km on average ($99$th percentile $4.1$--$10.1$~km; Table~\ref{tab:tau}, Figure~\ref{fig:tau}), approaching the direct LOS horizon over a single sol and, scaled diffusively over a $30$--$50$ sol storm, reaching tens of kilometres. That horizon is $2\sqrt{2Rh_a}\approx7.4$~km for two antennas at height $h_a=2$~m on a sphere model of Mars, i.e.,~radius $R=3390$~km.

The connectivity results of Section~\ref{sec:netresults} show that the swarm is held together not by a single long link but by multi-hop relay, and~with a wide margin: the critical range $r_c^\star=1.6$--$4.1$~km is a factor of $1.8$--$4.6$ below the $7.4$~km geometric horizon over one sol, and~the network is a redundant mesh. We frame this as a geometric connectivity requirement: the 7.4~km is a LOS horizon, not a demonstrated radio range. Whether a rolling, unstabilised, power-limited DR closes a few-kilometre link depends on antenna orientation, losses, and~a link budget beyond this paper's scope. Terrestrial long-range radio~\citep{foubert2020} establishes only that few-kilometre ranges are attainable in principle. Terrain occlusions between rovers would further reduce the effective range below the geometric horizon. Within~these caveats, the~analysis here indicates that a passive swarm meets the geometric connectivity requirement over  a sol with no active~station-keeping.

\subsection{Deployment, Non-Storm Conditions, and~Collisions}

While the DR swarm is designed to operate during a storm, little forecasting exists to guarantee that a landing coincides with storm onset, raising the question of how the swarm behaves before the storm arrives. A~GDS grows over days to weeks from a regional core before encircling the
planet~\citep{barnes2020,wolkenberg2020}, so its onset is detectable by existing orbiters well before it reaches a given site. The~rovers can therefore be held stowed and deflated---immobile, with~negligible aerodynamic cross-section---until a storm is detected, then inflated to deploy into the advancing front; an uninflated rover does not disperse. Quantifying non-storm dispersion directly---for instance, by repeating this analysis at the same solar longitude in a quiescent Mars year from the MCD and~testing how often the winds exceed the $9.4~\text{m}\,\text{s}^{-1}$ motion threshold---would improve the deployment-timing trade and is left to future~work.

Another consideration for deployment is the risk of collision between rovers. The~present model treats each rover as an independent point mass and does not resolve rover–rover contact. Because~neighbouring rovers share the large-scale wind effects, their relative velocities are set only by the gust differential. Simulated with spatially correlated gusts, the~relative speed between neighbouring rovers near deployment is $\sim$0.07~ms$^{-1}$ on average (95th percentile 0.36~ms$^{-1}$, maximum $\sim$2.1~ms$^{-1}$), giving impact energies below $\sim$0.7~J for 95\% of contacts, assuming a reduced mass of 10~kg (two 20~kg rovers). Such low-energy contacts are unlikely to threaten swarm cohesion; their effect on dispersal is not modelled here. Beyond~the first minutes, the~areal density of a $\leq$20-rover swarm is low enough that contacts are rare. To~further decrease the risks of collisions, a~staggered release with initial spacing exceeding one rover diameter at deployment could be utilised. Resolving contact dynamics explicitly is left to future~work.

\subsection{Generalising MY34 GDS to Other Storms and~Locations}
All the wind datasets and the resulting dispersion are derived from the MY34 GDS, specifically at the Curiosity landing site (Section~\ref{sec:wind}). Martian GDSs exhibit strong regional wind differences across latitudes, elevations, and~surface thermal inertia, so the specific separation values reported here should not be read as~universal.

Two considerations nonetheless support the broader relevance of the result. First, the~most recent planet-encircling storms share a common large-scale structure: Wolkenberg~et~al.~\citep{wolkenberg2020} found that the MY25, MY28, and MY34 events have a similar expansion-phase duration and that the two equinoctial storms (MY25 and MY34) encircled the planet by the same season ($L_s \approx 193^\circ$). The~near-surface wind envelope at the storm peak---the regime that drives the DRs---is therefore broadly representative across equinoctial GDSs rather than specific to~MY34.

Second, the~differences outlined in~\citep{wolkenberg2020} (e.g., the post-perihelion onset and longer decay of MY28, regionally distinct dust-lifting centres, and~the correlation of dust opacity with atmospheric and night-time surface temperatures) imply that the thermal forcing, \textls[-15]{and~hence, the surface winds, vary from site to site. These variations would shift the} mean wind and gust statistics, and~therefore the exact separation numbers, at~a different location. Hence, a~storm over a high-thermal-contrast region, or~wind channelling by local topography, could plausibly amplify~dispersion.

Crucially, the~connectivity conclusion does not rest on the precise wind values but on two site-independent facts: (i) storm-strength winds reach tens of m\,s$^{-1}$ in any GDS, well above the $\sim$9.4~m\,s$^{-1}$ threshold for DR motion (Section~\ref{sec:mobility}); (ii) that threshold, and~the hop-by-hop relay on which swarm connectivity depends (Section~\ref{sec:comm}), are properties of the vehicle and the swarm architecture, not of the observing site. Because~connectivity is maintained through nearest-neighbour relay rather than a single long link, a~regional increase in dispersion would enlarge the swarm's overall extent without necessarily breaking that connectivity. We therefore expect the qualitative conclusion that a passive, wind-driven swarm can remain connected over a multi-sol scale to~hold across sites. Repeating the analysis at a contrasting location---the northern-lowland Jezero crater, or~a southern-highland site---is a natural next step once the trajectory model is coupled to site-specific MCD~queries. 

\subsection{Limitations}

Three simplifications bound the scope of these results. First, the~terrain is taken as flat; real topography, including boulder fields, craters, and ridges~\citep{golombek2003}, would deflect trajectories and could either concentrate or disperse the swarm. Therefore, coupling the mobility envelope of Section~\ref{sec:mobility} to a digital elevation model is a clear next step. Previous work~\cite{doi:10.2514/1.A32132} explores terrain modelling for a similar rover construction, considering collisions with a randomised Martian rock field. While it demonstrates a tumbleweed's ability to traverse long-distances within a realistic Martian terrain model, it only focuses on a single rover. Additional work would be required to explore the dispersion and travel of a swarm within Mars' terrain. Second, the~wind is sampled from the storm-peak distribution and, for~the hour-long runs, from~its overall mean rather than each hourly regime. We overcome this limitation as the sol-length simulation uses the full diurnal cycle. Third, although~outputs are reported to metre precision for consistency with the MCD, these solutions do not imply metre-level accuracy: the MCD is itself a model constrained by limited remote sensing, without~a direct in situ ground truth for these fields~\citep{forget1999, Millour2018, Colatis2013}, so the true uncertainty is larger. None of these caveats overturns the central, order-of-magnitude finding that gust-driven dispersion remains well within the communication~horizon.

As we set out in Section~\ref{intro}, this study addresses swarm dispersion and connectivity, rather than vehicle subsystems, e.g.,~delivery and landing, power, thermal control, and~attitude control. Those decisions are investigated in other literature, e.g.,~\citep{antol2003, southard2007, forbes2010, ylikorpi2005,iglesias2020}. 

%=================================================================
\section{Conclusions}\label{sec:conclusions}

We have quantified the dispersion and communication connectivity of a passively driven tumbleweed rover swarm under Martian GDS conditions, combining a first-principles mobility analysis, a~multivariate wind model, and~deterministic and stochastic trajectory simulations. A~4~m, 20~kg DR is mobile in storm winds, clearing centimetre-scale obstacles and modest slopes. Building the wind model on the bivariate-normal distribution of the velocity components---rather than an independent Weibull in speed and heading---captures the true speed--direction dependence and the diurnal structure while keeping the stochastic dynamics analytically tractable through the OU process. The~deterministic and stochastic integrators describe the swarm dispersion: with a static rolling threshold enforced, two rovers separate by under 1~m over an hour (0.44~km over a sol) in the discrete white-gust case and by 6--25~m over an hour (1.4--3.3~km over a sol) for gust correlation times of 10--60~s. Modelled as a communication network, the~swarm requires only a 1.6--4.1~km per-rover range to remain connected over a sol (well within the 7.4~km geometric LOS horizon) and forms a mesh that stays connected across the plausible envelope---for gust amplitudes up to twice the nominal value with $\tau_g\le60$~s, and~at the nominal amplitude up to $\tau_g\approx150$~s (Table~\ref{tab:env})---and across gust spatial correlation lengths. A~heuristic $\sqrt{t}$ extrapolation indicates LOS relay could suffice for $\sim$3--21 sols depending on the gust correlation time. Overall, this study indicates that a passive swarm can meet the geometric connectivity requirement over a sol, subject to the assumed gust correlation time ($\tau_g \lesssim 60$~s), gust amplitude, unobstructed terrain, and~a radio link closing the few-kilometre critical~range.

Future work could build on these results by integrating surface topography into the trajectory model, calibrating the gust correlation time from sub-hourly wind measurements to exploit the full coloured-noise capability of the OU framework, adopting higher-order stochastic schemes such as the Milstein method, and~extending the propagation across the entire multi-sol life of a storm. Coupled with maturation of the DR flight \mbox{system---stowage}, power, thermal, and~payload---further development of our model would advance the tumbleweed swarm from a promising concept towards a deployable instrument for resilient, distributed Martian~\hl{science}%MDPI: Please check and confirm whether there are any Supplementary Materials associated with this manuscript. If yes, please add the section to the Back Matter and cite it in the main text. None.
.

%=================================================================
\vspace{6pt}

\authorcontributions{\textls[-15]{Conceptualisation, E.C.B. and G.M.T.D.; methodology, E.C.B.; software, E.C.B.; validation, E.C.B. and G.M.T.D.; formal analysis, E.C.B.; investigation, E.C.B.; \mbox{writing---original}} draft preparation, E.C.B.; writing---review and editing, E.C.B. and G.M.T.D.; visualisation, E.C.B.; supervision, G.M.T.D. All authors have read and agreed to the published version of the~manuscript.}

\funding{\hl{This} %MDPI: Information regarding the funder and the funding number should be provided. Please check the accuracy of funding data and any other information carefully. All correct.
 research was supported in part by a UKRI STFC DTP Studentship and an Oxford University Clarendon~Scholarship.}

\dataavailability{The wind datasets analysed in this study were retrieved from the Mars Climate Database (\url{https://www-mars.lmd.jussieu.fr/mars/access.html} (\hl{accessed on September 1 2026}%MDPI: Please provide the date you accessed the URL in the following format: “URL (accessed on Day Month Year)”.
)). The~simulation code and derived data generated for this study are available on the corresponding author's GitHub (@emmabelhadfa) \citep{Belhadfa2026-fq}.}

\acknowledgments{The authors thank \hl{Lisa Romkey} %MDPI: Titles (e.g., Dr., Mr., and Prof.) should NOT be used in the Acknowledgments section. We have removed them. Please confirm. Confirm
 for her support of this~work.}

\conflictsofinterest{The authors declare no conflicts of~interest.}

%=================================================================
\abbreviations{Abbreviations}{
The following abbreviations are used in this manuscript:\\

\noindent
\begin{tabular}{@{}p{0.8cm}l}
DR & Dust Rover\\
GDS & Global Dust Storm\\
\end{tabular}

\noindent
\begin{tabular}{@{}ll}
LOS & Line-Of-Sight\\
MCD & Mars Climate Database\\
MY & Mars Year\\
OU & Ornstein--Uhlenbeck\\
REMS & Rover Environmental Monitoring Station\\
RK4 & Fourth-Order Runge--Kutta\\
RTG & Radioisotope Thermoelectric Generator\\
\end{tabular}
}

%=================================================================
\begin{adjustwidth}{-\extralength}{0cm}
\reftitle{References}

%\bibliography{references}

\PublishersNote{}

\end{adjustwidth}
\end{document}